\documentclass[letter,11pt]{article}%

\usepackage{booktabs}
\usepackage{array, caption, threeparttable}
\usepackage[font=normalsize,labelfont=bf,labelsep=period]{caption}
\usepackage{indentfirst}
\usepackage[utf8]{inputenc}

\usepackage{geometry}
\usepackage{color}
\usepackage{amsmath}
\usepackage{amsthm}
\usepackage{amssymb}
\usepackage{graphicx}
\usepackage{caption}
\usepackage{setspace}
\usepackage{esint}
\usepackage{amsfonts}
\usepackage{epstopdf}%
\usepackage{booktabs}
\usepackage{epsfig}
\usepackage{lineno}
\usepackage{comment}
\usepackage{booktabs}
\usepackage{lscape}
\usepackage{array}
\usepackage[titletoc,title]{appendix}
\usepackage[labelsep=period]{caption}
\usepackage{epstopdf}
\usepackage{natbib}
\usepackage[colorlinks,citecolor=blue,urlcolor=blue,filecolor=blue]{hyperref}
\usepackage{diagbox}
\usepackage{mathtools}
\usepackage{threeparttable}
\usepackage{multirow}
\usepackage{float}
\usepackage{bm}
\usepackage{siunitx}
\usepackage{epsfig}
\usepackage{subcaption}
\usepackage{indentfirst}
\usepackage{array}
\usepackage{esint}
\usepackage{changepage}
\usepackage{array}
\usepackage{appendix}
\usepackage{xcolor}
\graphicspath{{figures/}}
\usepackage{makecell}  %表格换行
\usepackage{arydshln}  % 表格细线 \Xhline{0.4pt}  % 另一条较细的线
\usepackage{fullpage} % set 1 inch margins
\newcommand{\addtabletext}[1]{\raggedright #1}
\usepackage{ragged2e}
\usepackage{setspace} % double space
\providecommand{\U}[1]{\protect\rule{.1in}{.1in}}

\providecommand{\U}[1]{\protect\rule{.1in}{.1in}}
\providecommand{\U}[1]{\protect\rule{.1in}{.1in}}

\numberwithin{equation}{section}
\makeatletter
\theoremstyle{plain}

\theoremstyle{definition}

\theoremstyle{plain}

\RequirePackage[colorlinks,citecolor=blue,urlcolor=blue]{hyperref}
\RequirePackage{hypernat}
\providecommand{\U}[1]{\protect\rule{.1in}{.1in}}
\newtheorem{theorem}{Theorem}

\newtheorem{proposition}{Proposition}

\newtheorem{remark}{Remark}

\providecommand{\definitionname}{Definition}
\providecommand{\propositionname}{Proposition}
\providecommand{\theoremname}{Theorem}
\providecommand{\definitionname}{Definition}
\providecommand{\propositionname}{Proposition}
\providecommand{\theoremname}{Theorem}
\providecommand{\definitionname}{Definition}
\providecommand{\propositionname}{Proposition}
\providecommand{\theoremname}{Theorem}
\makeatother
\providecommand{\definitionname}{Definition}
\providecommand{\propositionname}{Proposition}
\providecommand{\theoremname}{Theorem}

\graphicspath{{figures/}}
\bibpunct[, ]{(}{)}{,}{a}{}{,}

\begin{document}
\captionsetup[figure]{labelfont={bf},labelformat={default},labelsep=period,name={Fig.}}
 
\title{A General CoVaR Based on Entropy Pooling \thanks{The authors are listed in alphabetical order; and all authors made equal contributions.}}
\author{Yuhong Xu\thanks{Corresponding author. Center for Financial Engineering, School of Mathematical Sciences, Soochow University, P. R. China. 
This work is supported by the Natural Science Foundation of China (No.12271391; No.11871050) and the Tang Scholar Fund.   Email: yhxu@suda.edu.cn}, \ \ \ Xinyao Zhao\thanks{ Center for Financial Engineering, School of Mathematical Sciences, Soochow University, P. R. China. }}

 \date{August 26, 2025}
 
\maketitle

\textbf{Abstract}: 
 
We propose a general CoVaR framework that extends the traditional CoVaR by incorporating diverse expert views and information, such as asset moment characteristics, quantile insights, and perspectives on the relative loss distribution between two assets.
To integrate these expert views effectively while minimizing deviations from the prior distribution, we employ the entropy pooling method to derive the posterior distribution, which in turn enables us to compute the general CoVaR. Assuming bivariate normal distributions, we derive its analytical expressions under various perspectives.  Sensitivity analysis reveals that CoVaR exhibits a linear relationship with both the expectations of the variables in the views and the differences in expectations between them. In contrast, CoVaR shows nonlinear dependencies with respect to the variance, quantiles, and correlation within these views.

Empirical analysis of the US banking system during the Federal Reserve's interest rate hikes demonstrates the effectiveness of the general CoVaR when expert views are appropriately specified. Furthermore, we extend this framework to the general $\Delta$CoVaR, which allows for the assessment of risk spillover effects from various perspectives.



\section{Introduction}
 
Since the 2008 financial crisis, systemic risk has garnered widespread attention, prompting the development of various measures and assessment methods. Among these, conditional value at risk (CoVaR), introduced by \cite{AdrianBrunnermeier2016}, has become one of the most widely used metrics. CoVaR quantifies the potential impact of distress in one asset on the stability of the entire financial system, capturing the comovement and contagion effects between financial assets. It provides a novel perspective for assessing systemic risk by evaluating the value at risk (VaR) of asset $Y$ (or the financial system) at a specified confidence level $\alpha$, conditional on asset $X$ experiencing extreme losses, i.e., $X = \text{VaR}_\alpha^X$. Subsequently, \cite{Girardi2013} extended this definition by modifying the condition to an inequality, $X \geq \text{VaR}_{\alpha}^X$, allowing for the consideration of more severe scenarios where asset losses exceed their VaR. Notably, CoVaR has significantly advanced systemic risk management and has been widely applied in research of risks faced by financial institutions \citep{Huang2016, Pellegrini2022, Torri2021}, as well as in energy markets \citep{Ji2019, Mensi2017}, currency markets \citep{Waltz2022, Zhao2023}, and other fields.

In practice, financial markets encompass a wide range of views and information that extend beyond the traditional condition of ${X=(\geq)\text{VaR}_\alpha^X}$. For instance, by utilizing the CME Group's FedWatch Tool \citep{CME}, we can estimate the probability distribution of future Federal Reserve interest rate hikes, a crucial factor in assessing systemic risk within the banking sector, particularly during periods of monetary tightening. Moreover, experts may hold specific views on key market factors such as the relative ranking, expected values, and volatility of asset losses. This raises an important question: how can we effectively measure asset risk while incorporating these information or expert views?

To address this challenge, we define a general CoVaR conditional on arbitrary information or expert views as: 
$\text{Pr}(Y \leq \text{CoVaR}_\alpha^{Y|V_{iew}}\mid V_{iew}) = \alpha$,
where $V_{iew}$ represents any views or information related to both $X$ and $Y$, or only $X$. This framework naturally includes the traditional cases where ${X=(\geq)\text{VaR}_\alpha^X}$.

So, how can general CoVaR be estimated? \cite{AdrianBrunnermeier2016} initially employed the quantile regression method to estimate CoVaR under the condition ${X=\text{VaR}_\alpha^X}$.  Later,  \cite{HuangLinHong2024}  developed a Monte-Carlo simulation-based batching estimator for CoVaR. Furthermore, the DCC-GARCH model and GARCH-copula model have been utilized to estimate CoVaR under conditions where ${X=(\geq)\text{VaR}_\alpha^X}$ \citep{Girardi2013, Karimalis2018}. 
However, these methods are insufficient when estimating the general CoVaR under views regarding the loss distribution or  moment characteristics of losses, such as $\tilde{\mu}_X \geq \tilde{\mu}_Y$, and $\tilde{\sigma}_X^2 \geq \sigma_Y^2$, where $\tilde{\mu}_X$, $\tilde{\mu}_Y$, $\tilde{\sigma}_X^2$, and $\tilde{\sigma}_Y^2$ represent perspective-based expectations and variances of $X$ and $Y$, respectively.

Fortunately, the entropy pooling (EP) approach provides a flexible theoretical framework that effectively integrates a variety of expert views \citep{Meucci2010}. The core idea behind EP approach is to convert these views into constraints and solve the associated optimization problem by ``distorting" the prior probability distribution under the minimum relative entropy criterion. This process generates an optimal posterior distribution that satisfies the specified views. Notably, a key advantage of the EP method is its ability to combine diverse types of views, including expectations, quantiles, volatilities, correlations, and rankings. For example, \cite{LiChen2024}, \cite{MbaAngaman2023}, and \cite{Meucci2014} applied the EP approach to portfolio selection, integrating views on expected returns, VaR, conditional volatility under GARCH models, and rankings. Moreover, the EP approach can be regarded as a generalized Bayesian method. Bayesian rules form the basis of the Black-Litterman model \citep{BL1992}, which is essentially a special case of the minimum relative entropy principle \citep{CatichaGiffin2006}. However, unlike the Black-Litterman model, which integrates views on asset returns assuming they follow a normal or elliptical distribution \citep{XiaoValdez2015}, the EP method is applicable to a wider range of market conditions, thus providing greater flexibility in modeling various risk perspectives.

In this article, we propose a general and flexible CoVaR framework that incorporates various views and information.  First, we show how to compute the general CoVaR numerically. 
To analytically study the general CoVaR, we assume that the joint prior and posterior distributions of two variables, $X$ and $Y$, follow bivariate normal distributions. Based on the EP method, we then derive the analytical expressions for the general CoVaR under various expert views. These views include expectations, variances, quantiles, the value of $X$, as well as correlation coefficients and relative views between $X$ and $Y$, both in equality and inequality forms.
Sensitivity analysis reveals that the general CoVaR exhibits linear relationships with the posterior expectation of $X$ and the difference in expectations between $X$ and $Y$, while its dependencies on variance, quantiles, and correlation are nonlinear.

Furthermore, we conduct an empirical analysis in the context of interest rate hikes and bank failures. We estimate the general CoVaR for several high-risk banks and the NASDAQ Bank Index under various views regarding rate hikes and  distressed banks. Notably, when the views on rate hikes are correctly specified, our CoVaR predictions often cover actual losses. However, under the scenario of SVB's losses, both traditional CoVaR and VaR tend to underestimate risk, while the quantile view performs best, as the CoVaR under this view covers the losses without significantly overestimating them.

Finally, we define the general $\Delta$CoVaR as the general CoVaR minus VaR, which is used to quantify risk spillover effects under various views. Under expectation-based and quantile views regarding SVB's losses, our $\Delta$CoVaR estimates for various banks usually exceed their losses, indicating that SVB exhibits high risk spillover effects under these views.

The remainder of this article is organized as follows. Section \ref{Sec 2} introduces CoVaR and the EP approach. In Section \ref{Sec 3}, we present the general CoVaR framework, including its definition, numerical implementation, and analytical expressions under the bivariate normal distribution. Furthermore, we define the general $\Delta$CoVaR along with its analytical expressions. Section \ref{Sec 4} presents empirical results for banks' CoVaR. Finally, the article is concluded in Section \ref{Sec 5}.

\section{Preliminaries}
\label{Sec 2}

\subsection{The traditional CoVaR}

VaR is one of the most widely used measures for evaluating financial risk. At a given confidence level $\alpha$, the VaR of asset $X$ is defined as $\text{Pr}(X\leq \text{VaR}_\alpha^X) =\alpha$, where $X$ denotes the loss of asset $X$. Therefore, $\text{VaR}_\alpha^X$  corresponds to the $\alpha$-quantile of the loss distribution of $X$. 
Building upon this concept, \cite{AdrianBrunnermeier2016} introduced the traditional CoVaR to capture systemic risk. Specifically, it measures the VaR of asset $Y$ at confidence level $\alpha$, conditional on asset $X$ experiencing a stress event, that is, when ${X = \text{VaR}_\alpha^X}$.
Mathematically, the traditional CoVaR is defined as: 
\begin{equation}
\begin{aligned}
 \text{Pr}(Y\leq \text{CoVaR}_\alpha^{Y|X=\text{VaR}_\alpha^X} \mid X=\text{VaR}_\alpha^X) =\alpha,
\end{aligned}
\end{equation}
assuming that the loss variables $X$ and $Y$ are continuous. 
It quantifies the potential impact of distress in asset $X$ on asset $Y$. 
While it can be estimated via quantile regression, this method is not valid for estimating our general CoVaR under most expert views. Fortunately, the entropy pooling approach can effectively integrate various forms of information and expert's opinions.

\subsection{Entropy pooling approach}

The entropy pooling (EP) approach, introduced by \cite{Meucci2010}, provides a powerful and general methodology for incorporating expert views into a prior probability distribution. This is achieved by minimizing the relative entropy, also known as the Kullback-Leibler (KL) divergence \citep{KL1951, Kullback}, between the posterior and prior distributions, subject to constraints that represent the expert views.
Formally, the relative entropy between a posterior distribution  $\tilde f(x)$  and a prior distribution $f(x)$ is defined as: \begin{equation} \varepsilon(\tilde{f}, f) = \int_{x \in \mathcal{X}} \tilde{f}(x) \left[ \ln \tilde{f}(x) - \ln f(x) \right] dx. \end{equation}
The EP approach seeks the posterior distribution $\tilde f$ that satisfies all the specified views while remaining as close as possible to the prior distribution in terms of relative entropy. One of the key advantages of this method is its ability to accommodate various types of views, including expectations, variances, quantiles, correlations, and even inequality constraints.

Given a view $V$ held with full confidence, the posterior probability density function $\tilde f_V$ is obtained by solving the following optimization problem:  
\begin{equation}
\tilde f_V=\underset{f \in \mathcal{V}}{\operatorname*{argmin}} \{\varepsilon(\tilde f, f)\}, 
\label{minf}
\end{equation}
where $\mathcal{V}$ represents the set of all distributions satisfying the given view $V$.

When incorporating multiple views $V_1, V_2, \dots, V_n$ with associated confidence levels $c_i \in (0,1)$ such that $\sum_{i=1}^n c_i = 1$, the final posterior distribution is:
\begin{equation}
\tilde f_V^* = \sum_{i=1}^n c_i \tilde f_{V_i},
\label{multiple views}
\end{equation}
where $\tilde f_{V_i}$ denotes the posterior probability density function under view $V_i$.

\section{A general CoVaR based on expert views}
\label{Sec 3} 
 
\subsection{A general framework}
 
We propose a general version of CoVaR conditional on expert views or information, denoted as $V_{iew}$, and defined as: 
\begin{equation}
\begin{aligned}
\textnormal{Pr}(Y \leq \text{CoVaR}_\alpha^{Y|V_{iew}} \mid V_{iew}) = \alpha.
\end{aligned}
\end{equation}
This metric evaluates the risk of asset $Y$ at confidence level $\alpha$, given the view $V_{iew}$. Here, ``expert views" refer to subjective assessments or predictions about future events, while ``information" encompasses objective elements such as historical events, verified facts, or current market observations. For brevity, we collectively refer to  expert opinions and information as ``views".

The condition $V_{iew}$ encompasses a broad range of views, significantly broadening the traditional CoVaR (which is conditioned on $X = \text{VaR}_\alpha^X$).
On one hand, $V_{iew}$ can represent various views regarding both $X$ and $Y$, such as their relative rankings and correlations. For instance:
\begin{quote}
(i) In a particular market environment, an expert may believe that the potential loss of one stock is lower than that 
of another.

(ii) Alternatively, drawing from the Black-Litterman model, $V_{iew}$ represents a relative view where the difference in losses between two assets follows a normal distribution.
 \end{quote}

On the other hand, $V_{iew}$ can also represent views solely on asset $X$, such as its marginal distribution, expectation, volatility, and quantiles. For example:
\begin{quote}
(iii) When assessing the risk of the US banking sector during the Federal Reserve's interest rate hikes, we use the CME Group's FedWatch tool to estimate the probability distribution of future rate changes. Then we can compute the risk under the view of the distribution type.
\end{quote}

Moreover, views can be formulated as equality or inequality constraints and may encompass multiple perspectives simultaneously, allowing for a highly flexible modeling of expert's views.

\subsection{Numerical implementation}
 
The general CoVaR typically lacks an analytical expression unless the underlying variables are assumed to follow certain classical distributions.  As a result, it is necessary to numerically solve the posterior distribution under various views to evaluate general CoVaR. Below, we outline a numerical implementation procedure based on the EP approach.

\textbf{Step 1:} Estimate the prior joint probability distribution. 
\begin{quote}
We estimate the joint distribution of asset losses $X$ and $Y$ using historical data. Appropriate methods include kernel density estimation, parametric distribution fitting, or copula-based techniques.
\end{quote}

\textbf{Step 2:} Derive the posterior joint probability distribution. 
 \begin{quote}
Using the EP approach, we compute the posterior joint probability distribution of asset losses $X$ and $Y$ conditional on the view  $V_{iew}$. Specifically, based on the principle of minimum relative entropy, we solve the following optimization problem using the Lagrangian dual method:  
\begin{equation}
\left\{
\begin{aligned}
\min_{\mathbf{\tilde{p}}} \quad & \varepsilon(\mathbf{\tilde{p}}, \mathbf{p}) = \sum_{j=1}^J \tilde{p}_j (\ln \tilde{p}_j - \ln p_j), \\
\text{s.t.} \quad & \mathbf{\underline{g}} \leq \mathbf{G} \mathbf{\tilde{p}} \leq \mathbf{\bar{g}},
\end{aligned}
\right.
\label{minf2}
\end{equation}
where $\mathbf{p}$ and  $\mathbf{\tilde{p}}$ are the prior and posterior joint probability distributions, each containing  $J$ elements, with $p_j, \tilde{p}_j > 0$, and satisfying $\sum_{j=1}^J p_j = 1$, and $\sum_{j=1}^J \tilde{p}_j = 1$. The view  $V_{iew}$ is incorporated through the linear constraint:  $\mathbf{\underline{g}} \leq \mathbf{G} \mathbf{\tilde{p}} \leq \mathbf{\bar{g}}$, where $\mathbf{G}$ is a $K \times J$ matrix, and $\mathbf{\underline{g}}$ and  $\mathbf{\bar{g}}$ are $K$-dimensional vectors. 
For instance, if the view specifies the expected value of $X$ is $\mu_1$, i.e., $\tilde{\mu}_X = \mu_1$, the corresponding constraints are $\mathbf{X}^\top \mathbf{\tilde{p}} = \mu_1$ and $\mathbf{1}^\top \mathbf{\tilde{p}} = 1$, where $\mathbf{X}$ denotes the loss of asset $X$, consisting of $J$ elements, and $\mathbf{1}$  represents a $J$-dimensional vector with each element equal to 1. In this case, we have $\mathbf{G}=\left(X, \mathbf{1}\right)^\top$ and $\mathbf{\underline{g}} = \mathbf{\bar{g}} = \left(\mu_1,1\right)^\top$.
\end{quote}

\textbf{Step 3:} Estimate the general CoVaR of $Y$ based on its marginal posterior distribution.
  
\vspace{5mm}
 
For continuous distributions, if the type of the joint posterior distribution is known, the optimal posterior parameters need to be determined to satisfy the given views while minimizing the relative entropy between the posterior and prior distributions. This requires solving the following optimization problem:
\begin{equation}
\left\{
\begin{aligned}
\min_{\mathbf{\tilde{\theta}}}  \quad & \varepsilon( f(x, y; \mathbf{\tilde{\theta}}), f(x, y; \mathbf{\theta})) =\int_{y \in \mathcal{Y}} \int_{x \in \mathcal{X}}  f(x, y; \mathbf{\tilde{\theta}}) \, \left [ \ln  f(x, y; \mathbf{\tilde{\theta}}) - \ln f(x, y; \mathbf{\theta})  \right ] \, dx\, dy,  \\
\text{s.t.} \quad & \mathbf{\underline{g}} \leq  \int_{y \in \mathcal{Y}}\int_{x \in \mathcal{X}}  g(x, y)f(x, y; \mathbf{\tilde{\theta}}) \, dx\, dy\leq \mathbf{\bar{g}}.
\end{aligned} 
\right.
\label{minf3}
\end{equation}
Here, $\tilde{\theta}$ and $\theta$ represent the parameters of the posterior distribution $f(x, y; \mathbf{\tilde{\theta}})$ and the prior distribution $f(x, y; \theta)$, respectively, and  $g(x,y)$ is a continuous function.  Methods such as neural networks, grid search, and other optimization techniques can be used to solve for the optimal parameters  $\tilde{\theta}$.

\begin{figure}[ht]
		\centering
   \begin{subfigure}[b]{0.45\textwidth}
    \centering
    \includegraphics[width=3.38 in]{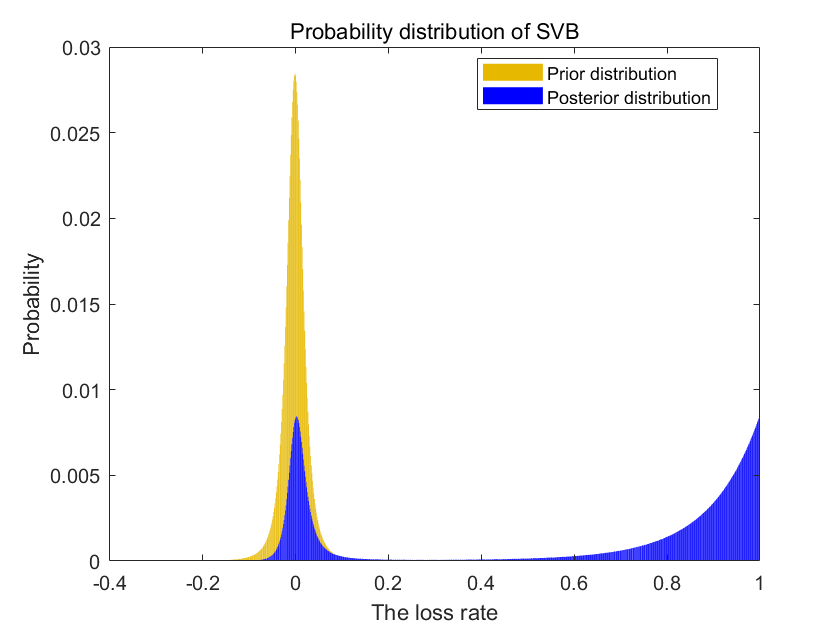}
    \caption{Prior and posterior distributions under the expectation view.}
  \end{subfigure}
  \hfill
  \begin{subfigure}[b]{0.45\textwidth}
    \centering
    \includegraphics[width=3.38 in]{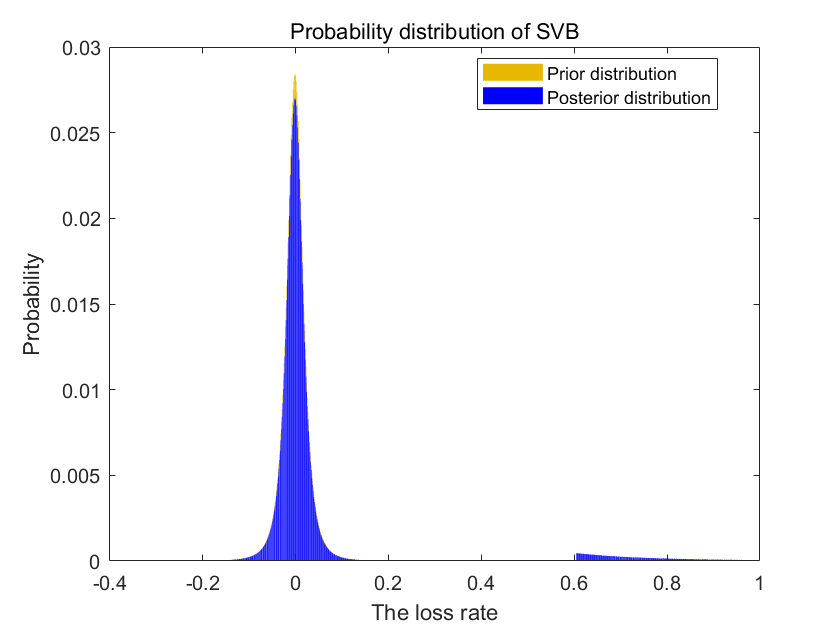}
    \caption{Prior and posterior distributions under the quantile view.}
  \end{subfigure}
\caption{Prior and posterior distributions without fixed types of the loss for SVB under the expectation view and the quantile view.}
\label{distribution_free}
\end{figure}

Through experimental analysis, we observed that if the posterior distribution's type is not specified, the resulting posterior distribution may become unreasonable. For example, we present the loss prior distribution of Silicon Valley Bank (SVB) and derive the corresponding posterior distributions under the expectation and quantile views, derived from Equation (\ref{minf2}). In Fig. \ref{distribution_free}\textcolor{red}{(a)}, the posterior distribution (represented in blue) exhibits a bimodal shape, with two distinct peaks far apart. Similarly, Fig. \ref{distribution_free}\textcolor{red}{(b)} shows a posterior distribution split into two distinct regions, with the probability in the middle approaching zero. These posterior distributions differ significantly from the typical distributions observed in financial data. To ensure that the results are meaningful and interpretable, we assume that the posterior distribution maintains the same distribution type as the prior distribution in subsequent empirical analyses.
  
\subsection{Analytical expressions under the normal distribution}
\label{Analytical expressions}

Now, we assume that the losses of two assets, $X$ and $Y$, follow a bivariate normal distribution, i.e., $(X,Y) \sim \mathcal{N}(\bm{\mu}, \mathbf{\Sigma})$, where $\bm{\mu} = \bm{\big(}\mu_X, \mu_Y \bm{\big)}^\top$ is the expectation vector, and  $\mathbf{\Sigma}=\begin{pmatrix}
\sigma_X^2 & \rho \sigma_X \sigma_Y \\ 
\rho \sigma_X \sigma_Y & \sigma_Y^2 
\end{pmatrix}$ is the covariance matrix, where 
$\sigma_X$ and $\sigma_Y$ are the standard deviations of $X$ and $Y$, and $\rho$ is the correlation coefficient between the two assets.
We further assume that the posterior distribution also follows a bivariate normal distribution with parameters:
$\bm{\tilde{\mu}} = \bm{\big(}\tilde \mu_X, \tilde \mu_Y\bm{\big)}^\top$ and $\mathbf{\tilde \Sigma} =\begin{pmatrix}  
\tilde \sigma_X^2 & \tilde \rho \tilde \sigma_X \tilde \sigma_Y \\ 
\tilde \rho \tilde \sigma_X \tilde \sigma_Y &\tilde \sigma_Y^2 
\end{pmatrix}$. 
The analytical expression for the relative entropy between these two bivariate normal distributions is as: 
\begin{equation}
\begin{aligned}
 \varepsilon(\tilde f, f)  = \frac{1}{2(1-\rho^2)}\left[\frac{(\tilde \mu_X-\mu_X)^2}{\sigma_X^2}-\frac{2\rho(\tilde \mu_X-\mu_X)(\tilde \mu_Y-\mu_Y)}{\sigma_X\sigma_Y}+\frac{(\tilde \mu_Y-\mu_Y)^2}{\sigma_Y^2}\right] \\
  +\frac{1}{2(1-\rho^2)}\left[\frac{\tilde \sigma_X^2}{\sigma_X^2}-\frac{2\rho \tilde \rho \tilde \sigma_X \tilde\sigma_Y }{\sigma_X\sigma_Y}+\frac{\tilde \sigma_Y^2}{\sigma_Y^2}\right]
-\frac{1}{2}\ln(\frac{1-\tilde \rho^2}{1-\rho^2})-\ln(\frac{\tilde \sigma_X \tilde\sigma_Y}{\sigma_X\sigma_Y})-1, \notag
\end{aligned}
\label{KL}
\end{equation}
where $\tilde f$ and $f$ represent the joint density functions of the posterior and prior distributions, respectively.

When the covariance matrix is non-singular, we can apply the Lagrangian method to solve the optimization problem in Equation (\ref{minf}) and determine the optimal posterior distribution $\tilde f$. Based on the analytical expression for the VaR of asset $Y$ under the normal distribution, 
$\text{VaR}^{Y}_\alpha=\mu_Y+\sigma_Y\Phi^{-1}(\alpha)$,
the CoVaR under the given view is expressed as:
\begin{equation}
\text{CoVaR}^{Y|V_{iew}}_\alpha = \tilde \mu_Y + \tilde \sigma_Y \Phi^{-1}(\alpha),
\end{equation} 
where $\Phi^{-1}(\alpha)$ denotes the inverse cumulative distribution function of the standard normal distribution. 
Thus, the key to deriving the analytical expression of CoVaR lies in determining the posterior expectation $\tilde \mu_Y$ and the posterior standard deviation $\tilde \sigma_Y$.

When the covariance matrix is singular, i.e., $\rho=\pm1$, the expressions for $\tilde \mu_Y$ and $\tilde \sigma_Y$ can be derived directly. Specifically, when $\rho=1$, the relationship between $X$ and $Y$ is perfectly positive linear, and $Y$ can be expressed as a deterministic function of $X$:
$Y=\frac{\sigma_Y}{\sigma_X}X+ \mu_Y-\frac{\sigma_Y}{\sigma_X} \mu_X$.
Similarly, when $\rho=-1$, the relationship between $X$ and $Y$ is perfectly negative linear, and $Y$ is also a deterministic function of $X$:
$Y=-\frac{\sigma_Y}{\sigma_X}X+\mu_Y+\frac{\sigma_Y}{\sigma_X} \mu_X$.\footnote{\ For a bivariate normal distribution, given $X = x$, the conditional distribution of $Y$ is expressed as:
$Y \mid X = x \sim \mathcal{N} \left( 
\mu_Y + \rho \frac{\sigma_Y}{\sigma_X}(x - \mu_X), \, (1 - \rho^2)\sigma_Y^2 
\right).$ When \(|\rho| = 1\), the conditional variance is zero, implying that $Y$ is a deterministic function of $X$, i.e., $Y= \mu_Y + \rho \frac{\sigma_Y}{\sigma_X}(X - \mu_X)$.} In such cases, any view expressed on $X$ is effectively equivalent to a view on $Y$.

Next, we introduce specific equality and inequality views and derive the corresponding analytical expressions for CoVaR.

\subsubsection{Views on expectation}
 
\label{ex_view}
If an expert believes that the expected loss of an asset in the future will reach (exceed or fall below) a certain level, then an expectation-type view can be incorporated into the analysis. This type of view essentially represents a statistical assessment of an asset's losses over a future horizon. If the expected loss reaches or exceeds a certain threshold, it implies that the risk level has surpassed a critical value. In such cases, it is meaningful to consider the asset's potential risk spillover effects on other assets.
Let \(\mu_1\) be a constant.  We consider three views regarding the expected loss of asset \(X\): \(\tilde{\mu}_X = \mu_1\), \(\tilde{\mu}_X \leq \mu_1\), and \(\tilde{\mu}_X \geq \mu_1\). These correspond to the scenarios where the posterior expected loss is equal to, less than, or greater than the constant \(\mu_1\).   
For these views, the corresponding CoVaR expressions for asset $Y$ are derived in the following theorem.

\begin{theorem}
Let $\mu_X$ and $\tilde \mu_X$ represent the prior and posterior expectations of $X$, respectively.

(i) Under the equality view  $\tilde \mu_X = \mu_1$, the \textnormal{CoVaR} for $Y$ is 
\begin{equation}
\textnormal{CoVaR}^{Y|\tilde \mu_X = \mu_1}_\alpha
= \mu_Y + \rho(\mu_1 - \mu_X)\frac{\sigma_Y}{\sigma_X} + \sigma_Y\Phi^{-1}(\alpha).
\label{normalmu}
\end{equation}

(ii) Under the inequality view $\tilde \mu_X \leq \mu_1$: 
\begin{quote} If furthermore $\mu_X \leq \mu_1$, then $\textnormal{CoVaR}_\alpha^{Y|\tilde \mu_X \leq \mu_1} =  \textnormal{VaR}^{Y}_\alpha$; 
\item  If furthermore $\mu_X > \mu_1$, then $\textnormal{CoVaR}_\alpha^{Y|\tilde \mu_X \leq \mu_1} =  \textnormal{CoVaR}_\alpha^{Y|\tilde \mu_X = \mu_1}$.  
\end{quote}

(iii) Under the inequality view $\tilde \mu_X \geq \mu_1$: 
\begin{quote} If furthermore  $\mu_X \geq  \mu_1$, then $ \textnormal{CoVaR}_\alpha^{Y|\tilde \mu_X \geq \mu_1} = \textnormal{VaR}^{Y}_\alpha$;  
\item  If furthermore $\mu_X < \mu_1$, then $ \textnormal{CoVaR}_\alpha^{Y|\tilde \mu_X \geq \mu_1}=  \textnormal{CoVaR}_\alpha^{Y|\tilde \mu_X = \mu_1}$.
\end{quote}
\label{thmu}
\end{theorem}

%The proof of Theorem \ref{thmu} is provided in \ref{proof1}\textcolor{blue}{.1}.

From Equation (\ref{normalmu}), if the losses of two assets are uncorrelated, imposing a constraint on the expected loss of asset $X$ will not affect the \textnormal{CoVaR} of $Y$, which will be equal to its VaR. However, if the losses of the two assets are positively correlated ($\rho > 0$), an increase in the expected loss $\mu_1$ of asset $X$ will lead to an increase in $\textnormal{CoVaR}_\alpha^{Y|\tilde \mu_X = \mu_1}$. Conversely, if $\rho < 0$, $\textnormal{CoVaR}_\alpha^{Y|\tilde \mu_X = \mu_1}$ will decrease as $\mu_1$ increases. When $\mu_1=\mu_X$, it indicates that the expert view coincides with the prior expectation and does not introduce any new information. Therefore, $\textnormal{CoVaR}_\alpha^Y$ is equal to  $\textnormal{VaR}_\alpha^Y$. Furthermore, $\textnormal{CoVaR}_\alpha^{Y|\tilde \mu_X = \mu_1}$ exhibits a linear relationship with the expert view on the expected loss of $X$.

The \textnormal{CoVaR} expressions under the inequality views can be intuitively explained. For example, in the first case of (ii), if both the expert view and the prior expectation are less than or equal to the constant $\mu_1$, i.e., $\tilde \mu_X \leq \mu_1$ and $\mu_X \leq \mu_1$, then the expert view does not introduce new information, and thus the CoVaR and VaR are the same. In the second case of (ii), when the expert view contradicts the existing facts  (i.e., \(\mu_X > \mu_1\)), the CoVaR value is determined by the intersection of the two intervals $\tilde \mu_X \leq \mu_1$ and $\mu_X > \mu_1$, or by the closest point between them, in accordance with relative entropy minimization.

\subsubsection{Views on variance}

Similarly, experts can also provide insights regarding the risk level  of an asset. For example, an expert might believe that the variance of asset $X$ will reach a specific level, denoted as $\tilde \sigma_X^{2} = \sigma_1^{2}$. Alternatively, experts might expect that the variance will either fall below or exceed a given threshold, represented as $\tilde \sigma_X^{2} \leq \sigma_1^{2}$ or $\tilde \sigma_X^{2} \geq \sigma_1^{2}$. The following theorem presents explicit expression for CoVaR under these different views on variance.
  
\begin{theorem}
Let $\sigma_X^2$ and $\tilde \sigma_X^2$ denote the prior and posterior variances of the loss for asset $X$, and let $\sigma_1$ be a positive constant.

(i) Under the equality view, where $\tilde \sigma_X^{2}=\sigma_1^{2}$, the $\textnormal{CoVaR}$ for $Y$ is 
 \begin{equation} 
 \begin{aligned} 
 \textnormal{CoVaR}_\alpha^{Y|\tilde \sigma_X^{2}=\sigma_1^{2}} = \mu_Y + \sigma_Y\left(1 - \rho^{2} + \rho^{2}\frac{ \sigma_1^{2}}{\sigma_X^{2}}\right)^\frac{1}{2} \Phi^{-1}(\alpha). 
 \end{aligned} 
 \label{normalsigma} 
 \end{equation}

(ii) Under the inequality view $\tilde \sigma_X^{2} \leq \sigma_1^{2}$: \begin{quote} If furthermore $\sigma_X^{2} \leq \sigma_1^{2}$, then $\textnormal{CoVaR}_\alpha^{Y|\tilde \sigma_X^{2} \leq \sigma_1^{2}} = \textnormal{VaR}_\alpha^{Y}$;
\item If furthermore $\sigma_X^{2} > \sigma_1^{2}$, then $\textnormal{CoVaR}_\alpha^{Y|\tilde \sigma_X^{2} \leq \sigma_1^{2}} = \textnormal{CoVaR}_\alpha^{Y|\tilde \sigma_X^{2} = \sigma_1^{2}}$. \end{quote}

(iii) Under the inequality view $\tilde \sigma_X^{2} \geq \sigma_1^{2}$: 
\begin{quote} If furthermore $\sigma_X^{2} \geq \sigma_1^{2}$, then $\textnormal{CoVaR}_\alpha^{Y|\tilde \sigma_X^{2} \geq \sigma_1^{2}} = \textnormal{VaR}_\alpha^{Y}$; 
 \item If furthermore $\sigma_X^{2} < \sigma_1^{2}$, then $\textnormal{CoVaR}_\alpha^{Y|\tilde \sigma_X^{2} \geq \sigma_1^{2}} = \textnormal{CoVaR}_\alpha^{Y|\tilde \sigma_X^{2} = \sigma_1^{2}}$. \end{quote} 
\label{thsigma} \end{theorem}

\begin{figure}[htbp]
    \centering    
    \includegraphics[width=0.5\linewidth]{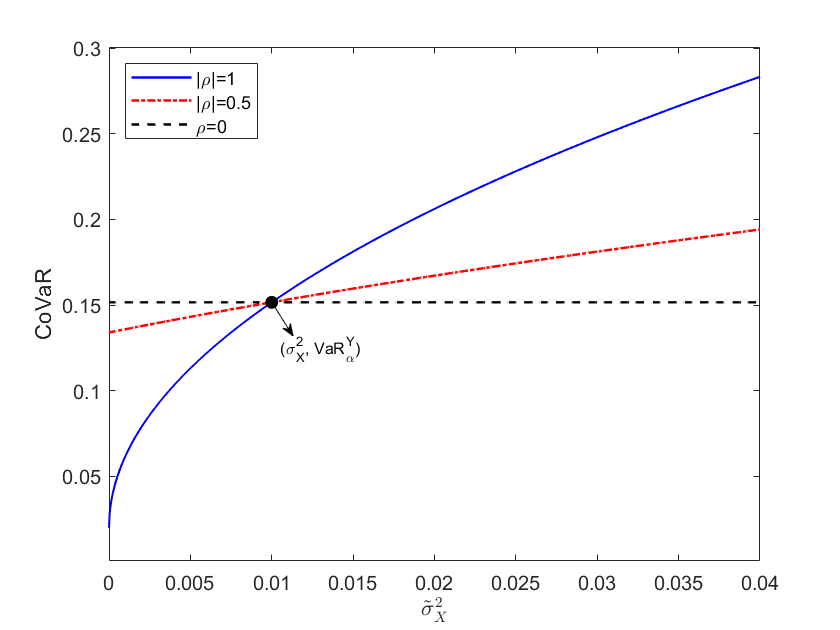}
    \caption{CoVaR for $Y$ under views on the variance of $X$. The random vector $(X,Y)$ follows a  bivariate normal distribution with parameters: $\mu_X = 0.10$, $\mu_Y = 0.02$, $\sigma_X^2 = 0.0100$, and $\sigma_Y^2 = 0.0064$. }
    \label{2 sigma}
\end{figure}

According to Equation (\ref{normalsigma}), we observe that when $\rho \neq 0$, the $\textnormal{CoVaR}_\alpha^{Y|\tilde \sigma_X^{2}=\sigma_1^{2}}$ exhibits a nonlinear and increasing relationship with the expert view on the variance of asset $X$'s loss. As illustrated in Fig. \ref{2 sigma}, when the expert assesses the variance to be greater than the prior variance, i.e., $\tilde \sigma_X^2 \geq \sigma_X^2$, the predicted $\textnormal{CoVaR}^Y_\alpha$ increases proportionally with the absolute value of the correlation coefficient. In this case, when $|\rho|=1$, the predicted  $\textnormal{CoVaR}^Y_\alpha$ reaches its maximum value.  Conversely, when the expert view suggests the variance is less than the prior variance, $\sigma_X^2$, the predicted $\textnormal{CoVaR}^Y_\alpha$ reaches its lowest value when $|\rho|=1$. These results indicate that a high correlation does not necessarily imply higher risk; instead, the predicted risk also depends on the expert view  regarding the variance. However, regardless of the value of $\rho$, when the  expert view on the variance coincides with the prior variance, i.e., $\tilde \sigma_X^{2}=\sigma_X^{2}$,  $\textnormal{CoVaR}$ equals its $\textnormal{VaR}$. This occurs because the expert view introduces no new information, causing the three curves to coincide at the point ($\sigma_X^{2}, \textnormal{VaR}_\alpha^Y$).

Similar to Section \ref{ex_view}, we can offer an intuitive explanation for the cases in (ii). In the first case of (ii), if both the expert view and the prior variance are below the constant $\sigma_1^2$, the expert view does not introduce any new information, resulting in CoVaR being equal to VaR. In the second case of (ii), when the expert view contradicts the existing facts, specifically when $\tilde \sigma_X^2 \leq \sigma_1^2$ but $\sigma_X^2 > \sigma_1^2$, the CoVaR value is determined by the intersection of the two intervals, or by the closest point between them.

\begin{remark}
If we have a combined view on both the expectation and variance, i.e.,
$\tilde \mu_X=\mu_1$ and $\tilde \sigma_X^{2}=\sigma_1^{2}$, the CoVaR expression for asset $Y$ is:
\begin{equation} 
\textnormal{CoVaR}^{Y|\tilde \mu_X =\mu_1, \ \tilde \sigma_X^{2}=\sigma_1^{2}}_\alpha
=\mu_Y+\rho(\mu_1-\mu_X)\frac{\sigma_Y}{\sigma_X}+   \sigma_Y\left(1 - \rho^{2} +\rho^{2}  \frac{\sigma_1^{2}}{\sigma_X^{2}}\right)^\frac{1}{2} \Phi^{-1}(\alpha). 
 \label{musigma}
\end{equation}
Moreover, we the following equality holds true:
\begin{equation} 
\textnormal{CoVaR}^{Y|\tilde \mu_X =\mu_1, \ \tilde\sigma_X^{2}=\sigma_1^{2}}_\alpha+\textnormal{VaR}^{Y}_\alpha=\textnormal{CoVaR}^{Y|\tilde \mu_X =\mu_1}_\alpha+  \textnormal{CoVaR}^{Y|\tilde \sigma_X^{2}=\sigma_1^{2}}_\alpha.
\label{equalityrelation}
\end{equation}
\end{remark}

\subsubsection{Views on quantile}

In the traditional CoVaR, the expert view is that the loss of asset $X$ equals its quantile, $X = q_X$. Here, the expert view does not specify the exact value of $X$, but instead takes the form $\tilde q_X = q_1$, $\tilde q_X \leq q_1$, and $\tilde q_X \geq q_1$, where $\tilde q_X$ and $q_X$ represent the posterior and prior $\alpha$ quantiles of $X$, respectively, and $q_1$ is a given constant.

\begin{theorem} 
(i) Under the equality view  $\tilde q_X=q_1$, the \textnormal{CoVaR} for $Y$ is expressed as
 \begin{equation} 
 \begin{aligned} \textnormal{CoVaR}^{Y|\tilde q_X=q_1}_\alpha
% &=\mu_Y+ \rho \sigma_Y \frac{q_1-q_X}{\sigma_X} +\left[\tilde\sigma_Y-\rho\left(\sigma_Y^2+\frac{\tilde\sigma_Y^2-\sigma_Y^2}{\rho^2}\right)^\frac{1}{2}+\rho\sigma_Y\right]\Phi^{-1}(\alpha),\\
 &=\left\{
\begin{aligned}
\mu_Y+ \rho (q_1-q_X) \frac{\sigma_Y}{\sigma_X} +\{\tilde\sigma_Y-\left[\tilde\sigma_Y^2-(1-\rho^2)\sigma_Y^2\right]^\frac{1}{2}+\rho\sigma_Y\}\Phi^{-1}(\alpha), \quad \text{if} \ \rho \geq 0,\\
\mu_Y+ \rho (q_1-q_X) \frac{\sigma_Y}{\sigma_X} +\{\tilde\sigma_Y+\left[\tilde\sigma_Y^2-(1-\rho^2)\sigma_Y^2\right]^\frac{1}{2}+\rho\sigma_Y\}\Phi^{-1}(\alpha), \quad \text{if} \ \rho<0,\\
%\mu_Y+  \sigma_Y\Phi^{-1}(\alpha), \quad \rho=0\\
\end{aligned} 
 \right.
 \end{aligned} \label{normalq} 
 \end{equation} 
 where $\tilde \sigma_Y=\sigma_Y\{\frac{1+(1-\rho^2)c^2}{1+c^2}+\frac{\rho^2(\kappa+c)c\left[(\kappa+c)c+\sqrt{(\kappa+c)^2c^2+4(1+c^2)}\right]}{2(1+c^2)^2}\}^\frac{1}{2}$, $\kappa= \frac{q_1- q_X}{\sigma_X}$, and $c=\Phi^{-1}(\alpha)$. 
\vspace{2mm}

Particularly, when $\rho=0$, $\textnormal{CoVaR}^{Y|\tilde q_X=q_1}_\alpha=\textnormal{VaR}_\alpha^Y$.
\vspace{2mm}
% \begin{equation} 
%\ \kappa= \frac{q_1- q_X}{\sigma_X}, \text{and} \ %c=\Phi^{-1}(\alpha).  \notag \end{equation} 

 %textcolor{red}{\rho\in [-1,0) \cup (0,1]}, 

(ii)  Under the inequality view $\tilde q_X \leq q_1$: 
 \begin{quote}  If furthermore $q_X \leq  q_1$, then $\textnormal{CoVaR}_\alpha^{Y|\tilde q_X\leq q_1}=\textnormal{VaR}^{Y}_\alpha$;
 \item If furthermore $q_X > q_1$, then $\textnormal{CoVaR}_\alpha^{Y|\tilde q_X\leq q_1}=\textnormal{CoVaR}_\alpha^{Y|\tilde q_X= q_1}$.  \end{quote}

(iii) Under the inequality view $\tilde q_X \geq q_1$:  
 \begin{quote}  If furthermore $q_X \geq  q_1$, then $\textnormal{CoVaR}_\alpha^{Y|\tilde q_X\geq q_1}=\textnormal{VaR}^{Y}_\alpha$;
\item If furthermore $q_X <  q_1$,  then $\textnormal{CoVaR}_\alpha^{Y|\tilde q_X\geq q_1}=\textnormal{CoVaR}_\alpha^{Y|\tilde q_X= q_1}$.  \end{quote} 
\label{thq} 
\end{theorem}

\begin{figure}[htbp]
    \centering
    \begin{subfigure}{0.5\textwidth}
        \centering
\includegraphics[width=\linewidth]{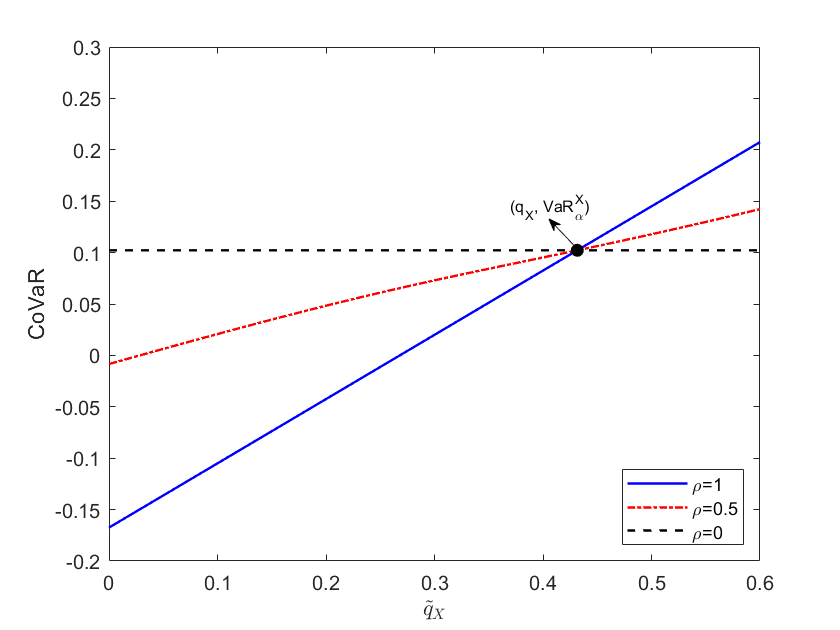}
  \subcaption{Positive correlation.}% coefficients.}
    \end{subfigure}%
    \hfill
    \begin{subfigure}{0.5\textwidth}
        \centering
        \includegraphics[width=\linewidth]{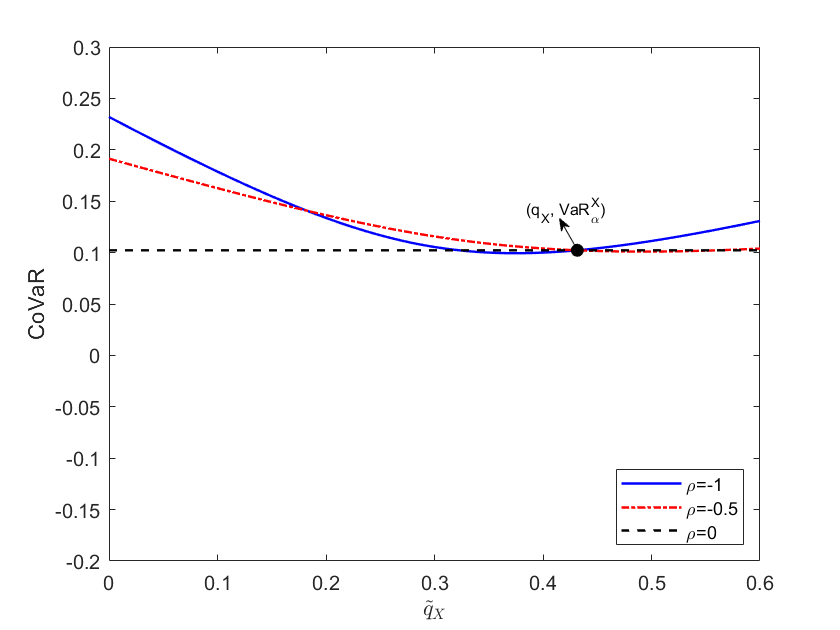}
    \subcaption{Negative correlation.}% coefficients.}
    \end{subfigure}
\caption{CoVaR for $Y$ under views on the 95\% quantile of $X$. The random vector $(X,Y)$ follows a  bivariate normal distribution with parameters: $\mu_X = 0.30$, $\mu_Y = 0.02$, $\sigma_X^2 = 0.0064$, and $\sigma_Y^2 = 0.0025$.}
\label{3 q}
\end{figure}

Next, we will analyze  $\textnormal{CoVaR}$ only under the equality view.
From the above theorem, we derive that when $\rho = 1$, $\textnormal{CoVaR}_\alpha^{Y|\tilde q_X=q_1} =\mu_Y+ (q_1-\mu_X)\frac{\sigma_Y }{\sigma_X}$, which exhibits a linear relationship with $q_1$. When  $0<\rho <1$, as shown in Equation (\ref{normalq}), $\textnormal{CoVaR}_\alpha^{Y|\tilde q_X=q_1}$  is nonlinear  with respect to $q_1$ and increases as $q_1$ increases. When $\rho=0$, the views on the quantile of $X$ have no effect on the CoVaR for $Y$, and $\textnormal{CoVaR}^{Y|\tilde q_X=q_1}_\alpha$ always equals  $\textnormal{VaR}_\alpha^Y$. However, when $-1\leq \rho < 0$, $\textnormal{CoVaR}_\alpha^{Y|\tilde q_X=q_1}$ exhibits a complex nonlinear relationship with $q_1$, as shown in Fig. \ref{3 q}.  
In both two subgraphs, when the expert view on the quantile coincides with the prior quantile $q_X$ of $X$'s loss, the three curves converges at a single point.

\subsubsection{Views on correlation coefficient}

Expert views on the correlation coefficient can also significantly influence the CoVaR. For example, during a financial crisis, a rational expert may believe that asset correlations will be equal to or exceed a certain level.
We discuss three views on the correlation coefficient: $\tilde \rho = \rho_1$, $\tilde \rho \leq \rho_1$, and $\tilde \rho \geq \rho_1$,  corresponding to cases where the posterior correlation coefficient equals, is lower than, or is higher than the constant $\rho_1$, respectively.

\begin{theorem}
Let $\rho$ and $\tilde \rho$ represent the prior and posterior correlation coefficients between assets $X$ and $Y$.

(i) Under the equality view $\tilde \rho=\rho_1$, the \textnormal{CoVaR} for $Y$ is
 \begin{equation} \begin{aligned} \textnormal{CoVaR}^{Y|\tilde \rho=\rho_1}_\alpha=\mu_Y+ \sigma_Y\left(\frac{1-\rho^2}{1-\rho\rho_1}\right)^\frac{1}{2} \Phi^{-1}(\alpha),\end{aligned} \label{normalrho} \end{equation}
where $\rho$ and $\rho_1$ both belong to $(-1, 1)$.\footnote{\ When $\rho=\rho_1=\pm 1$, $\textnormal{CoVaR}^{Y|\tilde \rho=\rho_1}_\alpha$ coincides with $\textnormal{VaR}^{Y}_\alpha$, since no additional information is introduced by  the view.  However, when $|\rho| = 1$ or $|\rho_1| = 1$ but $\rho \neq \rho_1$,  the relative entropy becomes infinite, indicating a significant divergence between the prior and posterior distributions.
To ensure continuity and maintain a well-defined framework,  we define the CoVaR as shown in Equation (\ref{normalrho}).}

(ii) Under the inequality view  $\tilde \rho\leq\rho_1$:   \begin{quote}  If furthermore $\rho\leq\rho_1$,  then $\textnormal{CoVaR}_\alpha^{Y|\tilde \rho \leq \rho_1}=\textnormal{VaR}^{Y}_\alpha$;
 \item If furthermore $\rho>\rho_1$,   then $\textnormal{CoVaR}_\alpha^{Y|\tilde \rho \leq \rho_1}=\textnormal{CoVaR}_\alpha^{Y|\tilde \rho = \rho_1}$. \end{quote}

 (iii)  Under the inequality view $\tilde \rho\geq\rho_1$: 
 \begin{quote}  If furthermore $\rho\geq\rho_1$, then
$\textnormal{CoVaR}_\alpha^{Y|\tilde \rho\geq \rho_1}=\textnormal{VaR}^{Y}_\alpha$;
 \item If furthermore $\rho<\rho_1$, then $\textnormal{CoVaR}_\alpha^{Y|\tilde \rho\geq \rho_1}=\textnormal{CoVaR}_\alpha^{Y|\tilde \rho = \rho_1}$. \end{quote} 
\label{thrho} \end{theorem}

\begin{figure}[htbp]
    \centering
    \includegraphics[width=0.5\linewidth]{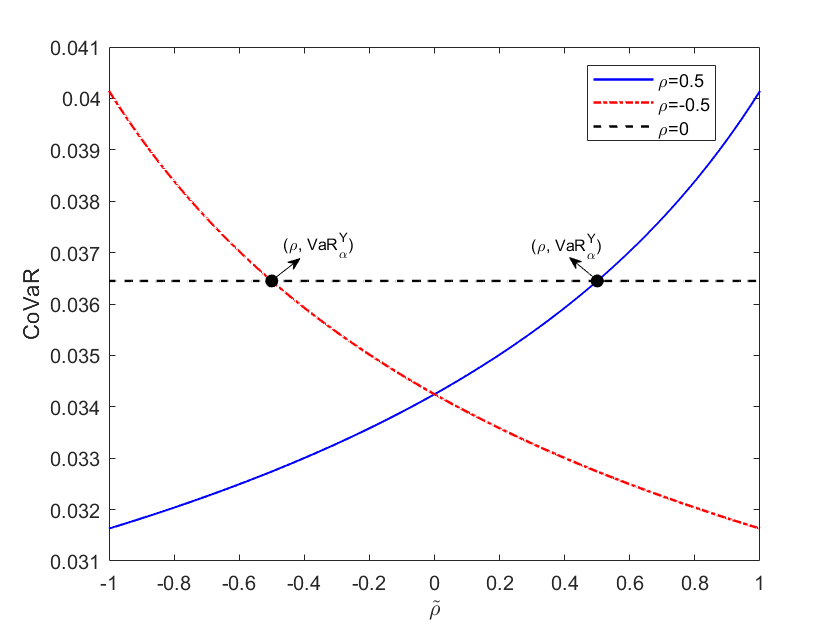}
    \caption{CoVaR for $Y$ under views on the correlation coefficient between $X$ and $Y$. The random vector $(X,Y)$ follows a  bivariate normal distribution with parameters: $\mu_X = 0.10$, $\mu_Y = 0.02$, $\sigma_X^2 = 0.0025$, and $\sigma_Y^2 = 0.0001$. }
    \label{4 rho}
\end{figure}

As illustrated in Fig. \ref{4 rho}, if the prior correlation coefficient is positive, i.e., $\rho>0$, $\textnormal{CoVaR}^{Y|\tilde \rho = \rho_1}_\alpha$ exhibits a nonlinear increasing relationship with the posterior correlation coefficient $\tilde \rho$. When $\tilde \rho>\rho>0$, $\textnormal{CoVaR}^{Y|\tilde \rho=\rho_1}_\alpha$  surpasses the VaR for $Y$. From a financial perspective, when the losses of two assets are positively correlated, a higher correlation between $X$ and $Y$ indicates a stronger relationship, resulting in a higher CoVaR for $Y$. 
Conversely, if $\rho<0$, the relationship exhibits the opposite monotonicity: $\textnormal{CoVaR}$ decreases as $\tilde \rho$ increases. Moreover, when $\tilde \rho$ is below $\rho$, $\textnormal{CoVaR}^{Y|\tilde \rho=\rho_1}_\alpha$ is higher than  VaR. 
Additionally, regardless of the value of $\rho$, when the posterior correlation coefficient coincides with the prior correlation coefficient, i.e., $\tilde \rho=\rho$, the CoVaR for $Y$ coincides with  the VaR for $Y$,  as indicated by the intersection points in Fig. \ref{4 rho}.

\subsubsection{Views on the value}
  
For views on the value of the loss of asset $X$: $X=l$, $X\geq l$, and $X\leq l$, the corresponding expressions for the \textnormal{CoVaR} of asset $Y$ are presented in Proposition \ref{propl}.

\begin{proposition}
Let $l$ be a constant, and $F(\cdot,\cdot)$ denote the cumulative distribution function of the bivariate normal distribution of $(X,Y)$.

(i)  Under the equality view $X=l$, the expression for \textnormal{CoVaR} of asset $Y$ is
\begin{equation}
\begin{aligned}
\textnormal{CoVaR}^{Y| X=l}_\alpha=\mu_Y+\rho(l-\mu_X)\frac{\sigma_Y}{\sigma_X}+\sigma_Y\left(1-\rho^{2}\right)^\frac{1}{2}\Phi^{-1}(\alpha).
\label{X=l}
\end{aligned}
\end{equation}

(ii)  When the inequality view is $X\leq l$, the \textnormal{CoVaR} is
$\textnormal{CoVaR}_\alpha^{Y| X\leq l} =F_{l,Y}^{-1}(\alpha\beta)$, where  $\beta=\textnormal{Pr}(X\leq l)=F_X(l)$. 

\vspace{3mm}

(iii) When the inequality view is $X\geq l$, the \textnormal{CoVaR} is $\textnormal{CoVaR}_\alpha^{Y| X\geq l}=F_{l,Y}^{-1}(F_{Y}(\textnormal{CoVaR}_\alpha^{Y| X \geq l}) -\alpha(1-\beta))$, where $F_{Y}(\cdot)$ is the marginal  distribution function of $Y$.
\label{propl}
\end{proposition}

%The proof of Proposition \ref{propl} is provided in \ref{proof1}\textcolor{blue}{.2}.
   
\begin{remark}
 
When $l=\textnormal{VaR}^X_\alpha$, the expression $\textnormal{CoVaR}^{Y| X=l}_\alpha$ corresponds to  the traditional \textnormal{CoVaR} proposed by \cite{AdrianBrunnermeier2016}.  
\end{remark}

\begin{remark}
If $l=\mu_1$, the view is $X=\mu_1$. In this case, the expression for the \textnormal{CoVaR} of asset $Y$ is
\begin{equation}
\begin{aligned}
\textnormal{CoVaR}^{Y| X=\mu_1}_\alpha=\mu_Y+\rho(\mu_1-\mu_X)\frac{\sigma_Y}{\sigma_X}+\left(1-\rho^{2}\right)^\frac{1}{2}\sigma_Y\Phi^{-1}(\alpha).
\label{X=l}
\end{aligned}
\end{equation}
Accordingly, the posterior expectation and variance of $Y$ are $\tilde \mu_Y=\mu_Y+\rho(\mu_1-\mu_X)\frac{\sigma_Y}{\sigma_X}$ and $\tilde \sigma_Y^2 =\left(1-\rho^{2}\right)\sigma_Y^2$, respectively.

By contrast, under the expectation view $\tilde \mu_X=\mu_1$, the \textnormal{CoVaR} of asset $Y$ is expressed as
\begin{equation}
\begin{aligned}
\textnormal{CoVaR}^{Y| \tilde \mu_X=\mu_1}_\alpha=\mu_Y+\rho(\mu_1-\mu_X)\frac{\sigma_Y}{\sigma_X}+\sigma_Y\Phi^{-1}(\alpha).
\label{X=l}
\end{aligned}
\end{equation}
In this case, $\tilde \mu_Y=\mu_Y+\rho(\mu_1-\mu_X)\frac{\sigma_Y}{\sigma_X}$ and $\tilde \sigma_Y^2 =\sigma_Y^2$.

Therefore, $\textnormal{CoVaR}^{Y| X=\mu_1}_\alpha \leq \textnormal{CoVaR}^{Y| \tilde \mu_X=\mu_1}_\alpha$, and the equality holds if and only if $\rho=0$. This is because when $\rho=0$, any view on $X$ has no effect on the \textnormal{CoVaR} of $Y$, in which case $\textnormal{CoVaR}^{Y|View}_\alpha = \textnormal{VaR}^{Y}_\alpha$.

Furthermore, as $|\rho|$ increases, the difference between $\textnormal{CoVaR}^{Y| \tilde \mu_X=\mu_1}_\alpha$ and $\textnormal{CoVaR}^{Y| X=\mu_1}_\alpha$ becomes larger. This is because $X=\mu_1$ implies that the posterior variance of $X$ is zero, and as the correlation between $X$ and $Y$ grows, the posterior variance of $Y$ decreases under view $X=\mu_1$, thereby amplifying the discrepancy.

\end{remark}

\subsubsection{Relative view of two variables}

The well-known Black-Litterman model incorporates a view in which the difference in losses between two assets, $(X - Y)$,  follows a normal distribution $\mathcal{N}(d, s^2)$, where $d$ is the posterior expectation and  $s^2$ is the posterior variance. This view is significant in finance because it enables investors to assess the expected relative performance between two assets, especially in markets with high uncertainty.  It helps to accurately evaluate risk and supports the optimization of portfolio management strategies.

\begin{theorem} 
Under the relative view where the difference between $X$ and $Y$ follows $\mathcal{N}(d, s^2)$, the CoVaR for $Y$ is
 \begin{align}
\textnormal{CoVaR}^{Y|(X-Y) \sim  \mathcal{N}(d, s^2)}_\alpha
&= \frac{\mu_Y \sigma_X (\sigma_X - \rho \sigma_Y) + (\mu_X - d) \sigma_Y (\sigma_Y - \rho \sigma_X)}{\sigma_X^2 - 2 \rho \sigma_X \sigma_Y + \sigma_Y^2} \notag \\
&+\sigma_Y\left(1+  \frac{\left[ s^2 - (\sigma_X^2 - 2 \rho \sigma_X \sigma_Y + \sigma_Y^2) \right] \left( \sigma_Y-  \rho \sigma_X \right)^2}{\left( \sigma_X^2 - 2 \rho \sigma_X \sigma_Y + \sigma_Y^2 \right)^2}\right)^{\frac{1}{2}}\Phi^{-1}(\alpha).
 \label{d1s1}
\end{align} 
\end{theorem}

As shown in Equation (\ref{d1s1}), the $\textnormal{CoVaR}^{Y|(X-Y) \sim  \mathcal{N}(d, s^2)}_\alpha$  exhibits a linear relationship with the expert's expected difference between the losses of the two assets, $d$. Specifically, if $\rho > \sigma_Y/\sigma_X$, the $\textnormal{CoVaR}$ increases as $d$ increases; if $\rho < \sigma_Y / \sigma_X$, $\textnormal{CoVaR}$ decreases as $d$ increases. 
In particular, if $\rho = \sigma_Y/\sigma_X$, $\textnormal{CoVaR}^{Y|(X-Y) \sim  \mathcal{N}(d, s^2)}_\alpha$ equals $\textnormal{VaR}^Y_\alpha$.  
Moreover, we observe that $\textnormal{CoVaR}$  shows a positive nonlinear relationship with the variance of the difference, $s^2$.

\subsection{A general \texorpdfstring{$\Delta$}{Delta}\text{CoVaR}}

\cite{AdrianBrunnermeier2016} defined $\Delta$CoVaR as follows:
\begin{equation} 
\Delta \textnormal{CoVaR}_\alpha^{Y|X}=  \textnormal{CoVaR}_\alpha^{Y|X=\textnormal{VaR}_\alpha^X} -
 \textnormal{CoVaR}_\alpha^{Y|X =\textnormal{VaR}_{50\%}^X},
 \label{trad_delta}
\end{equation}
which measures the difference in asset $Y$'s CoVaR when asset $X$ is under distress versus when $X$ is in its median state. This  $\Delta$CoVaR captures the increase in risk to asset $Y$ due to stress in asset $X$, reflecting the risk spillover from  asset $X$ to $Y$. It also captures the tail-dependency between two assets.

Unlike the traditional definition, we define a general $\Delta\textnormal{CoVaR}$ as the difference between our general CoVaR of asset $Y$ under a specified expert view, denoted $V_{iew}$, and the VaR of $Y$. This generalization enables us to quantify the incremental risk to asset $Y$ induced by incorporating specific expert views. The formal definition of $\Delta\textnormal{CoVaR}$ is given by: \begin{equation} \Delta\textnormal{CoVaR}_\alpha^{Y|V_{iew}} = \textnormal{CoVaR}_\alpha^{Y|V_{iew}} - \textnormal{VaR}_\alpha^{Y}. \end{equation} It is worth noting that our $\Delta\textnormal{CoVaR}$ can take either positive or negative values, depending on the specified view and the dependence structure between the losses of assets $X$ and $Y$.

Assuming both the prior and posterior distributions of $(X, Y)$ follow  bivariate normal distributions, the expression for $\Delta \textnormal{CoVaR}$ of asset $Y$ under an expectation view $\tilde{\mu}_X = \mu_1$ is:
\begin{equation}
\Delta\textnormal{CoVaR}^{Y|\tilde \mu_X = \mu_1}_\alpha
= \rho(\mu_1 - \mu_X)\frac{\sigma_Y}{\sigma_X}.
\label{normalmu1}
\end{equation} 
When $X$ and $Y$ are uncorrelated ($\rho = 0$), we have $\Delta \textnormal{CoVaR}_\alpha^{Y|\tilde{\mu}_X = \mu_1} = 0$, indicating no risk spillover effect from $X$ to $Y$. 
For $\rho > 0$, $\Delta\textnormal{CoVaR}^{Y|\tilde \mu_X = \mu_1}_\alpha$ increases as $\mu_1$ rises, 
implying that a higher expectation of asset $X$'s loss leads to greater risk spillover to asset $Y$.
Conversely, for  $\rho < 0$, $\Delta\textnormal{CoVaR}^{Y|\tilde \mu_X = \mu_1}_\alpha$ decreases as $\mu_1$ increases. When the posterior expectation of $X$ exceeds its prior expectation ($\mu_1>\mu_X$), indicating increased risk in $X$, and $\rho>0$, the resulting $\Delta\textnormal{CoVaR}^{Y|\tilde \mu_X = \mu_1}_\alpha$ is positive. In this case, the $\Delta\textnormal{CoVaR}^{Y|\tilde \mu_X = \mu_1}_\alpha$ is directly proportional to the correlation coefficient $\rho$ and the volatility ratio $\sigma_Y / \sigma_X$, which suggests that a higher correlation and a greater volatility ratio amplify the sensitivity of asset $Y$'s risk to changes in the expected loss of asset $X$.

Similarly, under a variance view where $\tilde{\sigma}_X^2 = \sigma_1^2$, the expression for $\Delta \textnormal{CoVaR}$ of asset $Y$ is:
\begin{equation}
\Delta\textnormal{CoVaR}^{Y|\tilde \sigma_X^{2}=\sigma_1^{2}}_\alpha
= \sigma_Y\left[\left(1 - \rho^{2} + \rho^{2} \frac{\sigma_1^{2}}{\sigma_X^{2}}\right)^\frac{1}{2} -1\right] \Phi^{-1}(\alpha).
\label{normalsigma1}
\end{equation} 
If the posterior variance of $X$ exceeds its prior variance and the two assets are correlated, the risk spillover effect from $X$ to $Y$ is positive. In this case, $\Delta\textnormal{CoVaR}^{Y|\tilde \sigma_X^{2}=\sigma_1^{2}}_\alpha$ is directly proportional to the absolute value of the correlation coefficient, the variance of $Y$, and the ratio of the posterior to the prior variance of $X$. 

For a combined view where both $\tilde{\mu}_X = \mu_1$ and $\tilde{\sigma}_X^2 = \sigma_1^2$, we derive the following relationship based on Equation (\ref{equalityrelation}) in Remark 3.1:
 \begin{equation} 
\Delta \textnormal{CoVaR}^{Y|\tilde \mu_X =\mu_1, \ \tilde \sigma_X^{2}=\sigma_1^{2}}_\alpha=\Delta \textnormal{CoVaR}^{Y|\tilde \mu_X =\mu_1}_\alpha+  \Delta \textnormal{CoVaR}^{Y|\tilde \sigma_X^{2}=\sigma_1^{2}}_\alpha.
\end{equation}
This decomposition implies that the total spillover effect from the combined view is the sum of the effects from the expectation view and the variance view.

Furthermore, under the condition $X=\textnormal{VaR}_\alpha^X$, our $\Delta\textnormal{CoVaR}$ for $Y$ is 
\begin{equation}
\begin{aligned}
\Delta\textnormal{CoVaR}^{Y|X=\textnormal{VaR}_\alpha^X}_\alpha= \sigma_Y\left[\left(1-\rho^{2}\right)^\frac{1}{2}-(1-\rho)\right]\Phi^{-1}(\alpha).\label{X=l1}
\end{aligned}
\end{equation}
If $X$ and $Y$ are unrelated, the risk spillover effect is zero. Furthermore, when $\rho> 0$, $\Delta\textnormal{CoVaR}^{Y|X=\textnormal{VaR}_\alpha^X}_\alpha$ is positive and directly proportional to $\sigma_Y$, while for $\rho< 0$, $\Delta\textnormal{CoVaR}^{Y|X=\textnormal{VaR}_\alpha^X}_\alpha$  is negative and  inverse proportional to  $\sigma_Y$.
When  $\rho \neq 0$, our general expression for $\Delta\textnormal{CoVaR}^{Y|X=\textnormal{VaR}_\alpha^X}_\alpha$  is lower than the traditional $\Delta$CoVaR, which equals $\rho \sigma_Y \Phi^{-1}(\alpha)$ under the bivariate normal assumption.
Moreover, $\Delta\textnormal{CoVaR}^{Y|X=\textnormal{VaR}_\alpha^X}_\alpha$ exhibits a complex nonlinear relationship with $\rho$.

The expressions for $\Delta$CoVaR under other views can be derived in a similar manner, which are shown in Appendix \ref{Appendix B}.

\section{Empirical analysis of the general CoVaR for US banks}
\label{Sec 4}
 
In this section, we estimate the general CoVaR under various views using empirical data  amidst Fed's rate hikes and recent bank failures, setting the confidence level at $\alpha=95\%$. First, we establish expert views on the effective federal funds rate during the Fed's 2022-2023 rate hikes and estimate the general CoVaR for Silicon Valley Bank (SVB). Second, we investigate various views related to SVB's collapse on March 10, 2023, assessing its impact on other banks and the NASDAQ Bank Index (NBI).
 
\subsection{The general CoVaR under views of rate hikes}

\subsubsection{Views of rate hikes}

Based on the methodology of the CME Group's FedWatch Tool, we calculate the market's daily opinions for future rate hikes.\footnote{\ The CME Group provides 30-day federal funds futures, the prices of which reflect the market's expectations for the average daily effective federal funds rate during the contract month. For more details, see \cite{CME}.}
Table \ref{view distribution} presents the average weekly probability distribution of four potential rate hike scenarios (25, 50, 75, and 100 basis points) observed one week prior to each rate hike announcement. Based on these distributions, we also compute the 95th percentile ($q_X$) and the expected rate hike ($erh$), and compare them with the actual rate hikes. The table covers the eight rate hikes that took place from March 2022 up to the collapse of Silicon Valley Bank (SVB). In most cases, the expected rate hikes closely matched the actual changes. However, a notable deviation occurred during the third hike on June 15, 2022, when the market anticipated a 54.30 basis point increase, considerably lower than the actual 75 basis points.
 
\begin{table}[htbp]
  \centering 
  \caption{Views of rate hikes and actual rate hikes (basis points). }
\renewcommand\arraystretch{1.2}
 % \resizebox{\textwidth}{!}{
    \begin{tabular}{cccccccccc}
    \toprule
    \multirow{2}{*}{Time}  &\multirow{2}{*}{\makecell{Date of\\  declaration}}   & \multicolumn{4}{c}{Distribution view} & \multirow{2}{*}{\makecell{View: \\ $X=\mathit{q_X}$}}& \multirow{2}{*}{\makecell{View: \\ $X=\mathit{erh}$}} & \multirow{2}{*}{\makecell{Actual\\ rate hike}}\\
 \cmidrule(l{2pt}r{2pt}){3-6}  
 &     & 25  &50  & 75  & 100 & &  &  \\
 \hline
    1   &  03/16/2022 & 0.9150  & 0.0850  & 0.0000  & 0.0000   & 50 & 27.13 & 25  \\
    2   & 05/04/2022  & 0.0234 & 0.9736  & 0.0030    & 0.0000  & 50  & 49.49& 50  \\
    3   & 06/15/2022  & 0.0000  & 0.8281  & 0.1719  & 0.0000 & 75  & \textbf{54.30} & 75 \\
    4   & 07/27/2022  & 0.0000  & 0.0000  & 0.7426  & 0.2574   & 100 & 81.43 & 75  \\
    5   & 09/21/2022  & 0.0000  & 0.0000  & 0.7857  & 0.2143   & 100 & 80.36  &  75 \\
    6   & 11/02/2022  & 0.0000  & 0.0848  & 0.9152  & 0.0000   & 75 & 72.88 & 75  \\
    7   & 12/14/2022   & 0.0000  & 0.7775  & 0.2225  & 0.0000  &75 & 55.56  & 50  \\
    8   & 02/01/2023   & 0.9630  & 0.0370  & 0.0000  & 0.0000  &25   & 25.93& 25  \\
%    9   & 03/22/2023 & 0.3208  & 0.6792  & 0.0000  & 0.0000  & 0.0000  & 16.98  & 25 \\
 %   10  & 05/03/2023 & 0.1833  & 0.8167  & 0.0000  & 0.0000  & 0.0000  & 20.42  &25 \\
        \bottomrule
    \end{tabular}%
 
\raggedright

  \vspace{1mm}
\addtabletext{\small{\quad\quad Notes: Columns 3 through 6  present the distribution view, detailing the \textbf{probabilities} assigned to the four rate hike scenarios (25, 50, 75, and 100 basis points). The final three columns report the 95th percentile, the expected rate hike, and the actual rate hike (basis points). }}
  \label{view distribution}
\end{table}% 

\subsubsection{The estimation of the general \text{CoVaR}}

Under the aforementioned rate hike scenarios, we estimate the general CoVaR for SVB to evaluate the systemic risk triggered by changes in monetary policy. The dataset spans from January 22, 1994, to February 1, 2023, and is sourced from \textit{wind.com.cn} and \textit{federalreserve.gov}. The in-sample data includes four historical rate hike periods: 1994-1995, 1999-2000, 2004-2006, and 2015-2018.  Based on these historical periods, we forecast the out-of-sample risk for the most recent rate hike cycle (2022-2023). To mitigate the effects of stock price volatility and more accurately capture the impact of interest rate hikes on stock losses, we use weekly average prices.\footnote{\ Since most rate hikes do not take effect on Mondays, we define each week as a seven-day period starting the day before the rate hike, ensuring consistency in the weekly rates. We then calculate the weekly averages for each variable. The confidence level for CoVaR is set at 95\%.}

We first use the quantile regression (QR) method to estimate the traditional CoVaR under the view ${X=q_X}$. 
Given the view that the effective federal funds rate hike equals its VaR at the $\alpha$-confidence level (i.e., $X=q_X$), the CoVaR of SVB is calculated as:
\begin{equation}
\begin{aligned}
\textnormal{CoVaR}^{Y|X=q_X}_t(\alpha)=\hat\beta_0^{Y|X}(\alpha) + \hat\beta_1^{Y|X}(\alpha) q_X, \notag
\end{aligned}
\end{equation}
where $\hat\beta_0^{Y|X}(\alpha)$ and $\hat\beta_1^{Y|X}(\alpha)$ are the estimated regression coefficients at the $\alpha$-confidence level.

Table \ref{SIVB interest EP Expection} reports the CoVaR estimates for SVB under various interest rate hike scenarios. The CoVaR estimates based on the QR method under the $X=q_X$ view generally cover the actual losses, except in the case of the third hike. This suggests that, overall, the QR method is effective in forecasting SVB's tail risk during periods of rate hikes.

\begin{table}[htbp]
\centering
 \caption{The general CoVaR of SVB under eight-time interest hikes using QR and EP methods.}
    \renewcommand\arraystretch{1.2}
  \resizebox{\textwidth}{!}{
\begin{tabular}{ccrc@{\hspace{20pt}} c@{\hskip 2em} ccc}
        \toprule 
    \multirow{2}{*}{Time} & \multicolumn{1}{c}{\multirow{2}[0]{*}{Date}}     & \multicolumn{1}{c}{\multirow{2}[0]{*}{Loss  (\%)}}   & \multicolumn{1}{c}{CoVaR (\%): QR} & & \multicolumn{3}{c}{CoVaR (\%): EP}\\
   \cmidrule{6-8}   
 &     &   & \multicolumn{1}{c}{$X=q_X$}  & & \multicolumn{1}{c}{$X=q_X$}  & \multicolumn{1}{c}{$X=erh$} & \multicolumn{1}{c}{Distribution} \\
     \midrule
  1 & 03/17/2022   & -6.96   &7.1  &    & 6.7  & 6.4  & 6.4   \\
  2 & 05/05/2022    & 7.00   & 7.0  &    &  \textbf{6.4} &  \textbf{6.4} &\textbf{6.4}    \\
  3 & 06/16/2022   & 7.33    & \textbf{7.3}  &   &\textbf{7.3}   & \textbf{7.3}    &\textbf{7.0} \\
  4 & 07/28/2022   & -3.70   &7.5  & &  7.6   & 7.6    &7.6 \\
  5 & 09/22/2022   & 7.67    &7.9   &   & 7.9  & \textbf{7.3}    &7.9\\
  6 & 11/03/2022   & 6.93    &7.7 &  & 8.5   & 8.5   & 8.5  \\
  7 & 12/15/2022   & 5.16    &7.5 & & 8.8  &  8.5     & 8.2 \\
  8 & 02/02/2023    & -6.96  &7.2   &  & 7.3   & 7.3 &  7.3  \\
%\hline
%  Kupiec test& &      &Accept &  &Accept  &Reject &Accept  \\
\hline
\multirow{2}{*}{\makecell{Mean  of \\ absolute differences}}& &    	& \multirow{2}{*}{5.4} &  	& \multirow{2}{*}{5.6} 	& \multirow{2}{*}{5.7}  & \multirow{2}{*}{5.6} 		 \\
 
  &&&&&&\\
\multirow{2}{*}{\makecell{Variance of \\ absolute differences}} & &  	& \multirow{2}{*}{43.1}  &  &  \multirow{2}{*}{39.0} 	& \multirow{2}{*}{39.6}	& \multirow{2}{*}{38.9} \\
  &&&&&&\\
  \bottomrule  
  \end{tabular}%
   }
\raggedright

  \vspace{1mm}
\addtabletext{\small{\quad\quad Notes: The term ``absolute difference" refers to the absolute value of the difference between the CoVaR estimate and the actual loss. }}
 
 \label{SIVB interest EP Expection}
 \end{table}

However, to fully incorporate market views on interest rate hikes, especially the information contained in their probability distribution, we employ the EP approach,  as shown in Equation (\ref{multiple views}). Given the heavy-tailed nature of the data, we fit the marginal distributions using the $t$-distribution, which is more suitable for capturing extreme financial markets events. Furthermore, since the dependence between financial variables tends to intensify under extreme conditions, we model the dependence structure using a  $t$-copula. This method allows for a more accurate estimation of the prior joint distribution of bank losses and interest rate hikes. We then apply the EP approach to derive the posterior joint distribution and the corresponding CoVaR based on three distinct views:  $X=q_X$, $X=\mathit{erh}$, and the probability distribution of rate hikes,  as detailed in Table \ref{view distribution}. According to Table \ref{SIVB interest EP Expection}, under the EP-based $X=q_X$ and probability distribution views, there are two exceptions where the CoVaR estimates fail to cover the actual losses. Under the  $X=\mathit{erh}$ view, three exceptions are observed. Therefore, expectation-based views cannot capture extreme market conditions.

To further compare the performance of different general CoVaR models, we compute the mean and variance of the absolute  differences between the CoVaR estimates and actual losses, as shown in the last two rows of Table \ref{SIVB interest EP Expection}. The CoVaR estimates based on the QR method exhibit the lowest average absolute difference. This result is primarily due to the smoothing effect of the QR method, which reduces the sensitivity of CoVaR to extreme values. However, the CoVaR estimates obtained using the EP approach demonstrate better stability, as indicated by their lower variance in the absolute differences. Specifically, the general CoVaR under the distribution-based view shows the lowest volatility, as this view incorporates more comprehensive market information compared with moment- or quantile-based views. In addition, we estimate the general CoVaR for the NBI index to assess the performance of CoVaR models across various assets under different rate hike scenarios (see Appendix \ref{Robustness under views of interest hikes}). Our results indicate that under the probability distribution view, the CoVaR estimates exhibit both the lowest mean and variance of the absolute differences between CoVaR estimates and actual losses, highlighting its advantages in terms of predictive accuracy and stability.

It is worth emphasizing that the effectiveness of CoVaR estimation largely depends on the accuracy of the specified view. For example, during the third rate hike on June 16, 2022, a significant deviation between market expectations and the actual rate hike led to the underestimation of actual losses.
Moreover, the key strength of the EP approach lies in its ability to incorporate diverse types of views. For instance, integrating market beliefs about the probability distribution of rate hikes leads to improved predictive performance. Other types of views are explored in the following subsections.

%%%改到这里了
\subsection{The CoVaR and \texorpdfstring{$\Delta$}{Delta}\text{CoVaR} for banks and the NBI under various views on SVB}
\label{CoVaR SVB}

During the Fed's interest rate hike cycle, SVB filed for bankruptcy on March 10, 2023. In the aftermath of its failure, several other banks experienced significant losses, including Signature Bank (SB), First Republic Bank (FRB), First Horizon Bank (FHN), Western Alliance Bank (WAL), and PacWest Bank (PACW).  Among them, SB and FRB subsequently failed on March 12 and May 1, respectively. Against this backdrop, we assess the risk faced by these five banks, as well as the NASDAQ Bank Index (NBI), under the condition of a sharp decline in SVB's stock price.

The dataset used in this analysis contains the daily loss rates of these six banks and the NBI, spanning from January 1, 2015, to the day before SVB's failure. Table \ref{Correlation} reports the correlation coefficients between SVB and the other banks and the NBI. The results show that all correlation coefficients exceed 0.5 and are statistically significant at the 1\% level, indicating a strong positive correlation in losses between SVB and the other banks.

\begin{table}[htbp]
		\centering
\caption{Correlation coefficients between the losses of SVB and other banks or the NBI.}
\renewcommand\arraystretch{1.2}
    \begin{tabular}{ccccccc}
    \toprule
  &SB   &FRB  &FHN  &WAL  &PACW  &NBI\\
    \hline 
 SVB  &0.67 &0.72 &0.58	&0.72 &0.72	&0.75\\

 % FRB   &/ &/ &0.47	& 0.63	& 0.57	& 0.54\\
 \bottomrule
    \end{tabular}%
   \label{Correlation}%
\end{table}%

\subsubsection{Comparative analysis of the \text{CoVaR} under various types of views}

We consider five types of views on SVB's stock loss rate, denoted as $X$: the expectation view, the joint expectation and variance view, the quantile view, the traditional CoVaR condition, and the no-view case (i.e., without additional information).  A detailed description of these views is provided in Table \ref{Views Description}.

\begin{table}[htbp]
  \centering
  \caption{Views description.}
  \renewcommand\arraystretch{1.2}
    \centering
 
     \begin{tabular}{ll}
  \toprule
Views & Description\\
  \hline
%(a) $X=l$& The value of the loss rate equals $l$\\  
(a) $\tilde \mu_X=l$ & Expectation view: The expected loss rate equals $l$.\\  
\multirow{2}{*}{(b) $\tilde \mu_X=l, \tilde \sigma_X^2=\sigma_1^2$} &Expectation and variance view: The expected loss rate equals $l$, \\
& and the variance equals $\sigma_1^2$.\\  
(c) $\tilde q_X=l$ & Quantile view: The 95\% quantile of the loss rate equals $l$.\\ 
(d) $X=q_X$ &  Traditional CoVaR condition: The loss rate equals its  VaR$_{95\%}^X$ ($q_X$). \\
(e) No view & In this case, the CoVaR equals the VaR.\\
  \bottomrule
    \end{tabular}
 
\raggedright
  \vspace{1mm}
\addtabletext{\small{\quad\quad Notes: $l$, $\sigma_1$, and $q_X$ are specified constants.}}
  \label{Views Description} 
\end{table} 

In our empirical analysis, we set the loss level $l$ for views (a), (b), and (c) to SVB's loss rate on the day before its failure, which is 60.41\%. In view (b), $\sigma_1^2$ represents the variance of loss rates over the 10 days preceding to failure, estimated at 3.60\%. The 95th percentile of SVB's loss rate, denoted as $q_X$, is 4.24\%. Based on these parameters, we estimate the CoVaR of the banks on the day of SVB's collapse.

We have demonstrated in Section \ref{Analytical expressions} that CoVaR exhibits monotonicity with respect to the parameters specified in different views under the assumption of a bivariate normal distribution. Given the positive correlation between bank loss rates and the parameter setting $l > q_X$, the following conclusions can be drawn:
\begin{quote} 
(i) The CoVaR under views (a) and (b) is generally higher than that under view (c). First, the CoVaR under views (a) and (b) exceeds the CoVaR corresponding to the condition $X = l$, since $X = l$ is equivalent to setting $\tilde{\mu}_X = l$ and $\tilde{\sigma}^2 = 0$, whereas $\tilde{\sigma}^2 > 0$ under views (a) and (b). Second, the condition $X = l$ itself represents a more extreme scenario than $\tilde{q}_X = l$. Hence, the CoVaR under $X = l$ is higher than that under $\tilde{q}_X = l$.

(ii) Similarly, the CoVaR under views (a) and (b) is also higher than that under view (d).

(iii) The CoVaR under view (c) is higher than that under view (d), since in view (d), $\tilde{q}_X = q_X < l$. 
\end{quote}
In the context of bank failures,  CoVaR under view (d), i.e., traditional CoVaR, is higher than VaR (under view (e)).  In summary, the CoVaR  estimates various different views follow the ranking: (a) or (b), (c), (d), (e).

On the day of SVB's failure, we estimate the general CoVaR for the five banks and the NBI based on the five views described above. The results, presented in Table \ref{SVB CoVaR Copula Fixed}, confirm this ranking. Under the expectation-based views (a) and (b), the CoVaR estimates not only exceed the actual loss rates but also surpass the largest daily losses observed during the three days following SVB's failure. This is mainly due to the extreme assumptions in these views and the heavy-tailed property of the $t$-distribution, which is effective in capturing extreme events. Furthermore, view (a) only alters the prior expectation of SVB's loss rate without imposing any constraint on its variance. Therefore, we assume $\tilde \sigma_X^2 = \sigma_X^2$, which is significantly lower than the variance specified in view (b), $\tilde \sigma_X^2=3.60\%$. As a result, the CoVaR under view (a) is lower than that under view (b). Notably, under extreme conditions, the CoVaR estimated based on this expectation view exhibits the strongest risk amplification effect. The third-highest CoVaR arises from the quantile view (c), whose estimate is relatively close to the actual loss. In contrast, the traditional CoVaR (view (d)) fails to cover the actual losses for SB, FRB, WAL, and PACW, indicating its limitations in capturing extreme events. View (e), which represents the baseline case with no additional information and corresponds to the 95\% VaR, consistently yields the lowest CoVaR values.  To sum up, under extreme market conditions, traditional CoVaR and VaR  tend to underestimate risk, particularly the losses observed in the three days following SVB's failure. In comparison, the quantile view (c) performs best, as its corresponding CoVaR covers the losses without significantly overstating the risk.

\begin{table}[ht]
  \centering
  \caption{The general CoVaR of five banks and NBI under various views on SVB.}
  \renewcommand\arraystretch{1.3}
 \resizebox{\textwidth}{!}{
     \begin{tabular}{llrrrrrr}
  \toprule
&Views  &SB  &FRB   &FHN &WAL &PACW &NBI \\
    \hline
Loss (\%) &\ / & 22.87 & 14.84 & 3.97 & 20.88 & 37.91 & 4.64 \\
Worst loss (\%)&\ /   & 100.00 & 61.83 & 20.20 & 47.06 & 37.91 & 26.05\\
[2mm]
\hline
\multirow{7}{*}{CoVaR (\%)}%&(a) $X=60.41\%$    &  68.22  & 45.40  & 79.00  & 63.46  & 68.08  & 44.28  &Accept\\
&(a) Expectation view $\tilde \mu_X=60.41\%$ & 71.44  & 55.20  & 68.78  & 59.12  & 61.22  & 49.88 \\
& \multirow{2}{*}{\makecell{(b) Expectation and variance view  \\ [1mm] \hspace{2mm}$\tilde \mu_X=60.41\%, \tilde \sigma_X^2=3.60\%$}} & \multirow{2}{*}{74.38} & \multirow{2}{*}{74.10} & \multirow{2}{*}{74.38} & \multirow{2}{*}{75.64} & \multirow{2}{*}{74.24} & \multirow{2}{*}{75.64}  \\ 
& &&&&&& \\
&(c) Quantile view $\tilde q_X=60.41\%$ & 34.90  & 29.72  & 30.70  & 34.48  & \textbf{34.76} & 34.34 \\
& \multirow{2}{*}{\makecell{(d) Traditional CoVaR condition\\ [1mm] 
\hspace{2mm}$X=q_X=4.24\%$ }} & \multirow{2}{*}{\textbf{4.52}} & \multirow{2}{*}{\textbf{4.80}} & \multirow{2}{*}{5.64} & \multirow{2}{*}{\textbf{6.48}} & \multirow{2}{*}{\textbf{6.48}} & \multirow{2}{*}{4.80}  \\ 
& &&&&&& \\
 & (e) No view   &\textbf{3.68}&\textbf{2.70} & \textbf{2.98} &\textbf{3.54}  & \textbf{3.68}& \textbf{2.56} \\
   \bottomrule
    \end{tabular}%
   }
   
\raggedright

\vspace{1mm}
\addtabletext{\small{\quad\quad Notes: The ``loss" refers to the loss experienced by the corresponding bank on the day of SVB's failure, while the ``worst loss'' denotes the largest single-day loss incurred by the corresponding bank within the three days following SVB's failure.}}
  \label{SVB CoVaR Copula Fixed}%
\end{table}%

\begin{figure}[htbp]
		\centering
   \begin{subfigure}[b]{0.45\textwidth}
    \centering
    \includegraphics[width=3.38 in]{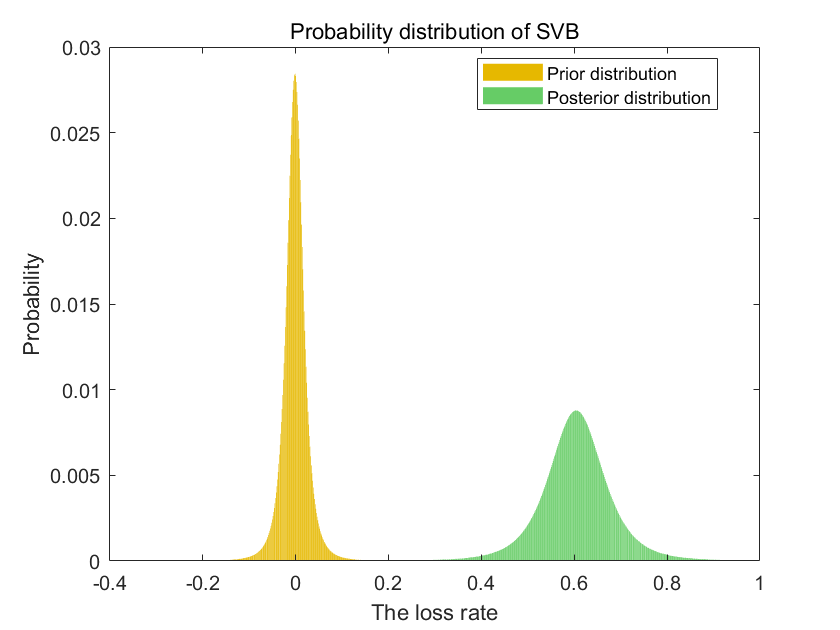}
    \caption{Prior and posterior distributions of loss for SVB under the expectation view.}
  \end{subfigure}
  \hfill
  \begin{subfigure}[b]{0.45\textwidth}
    \centering
    \includegraphics[width=3.38 in]{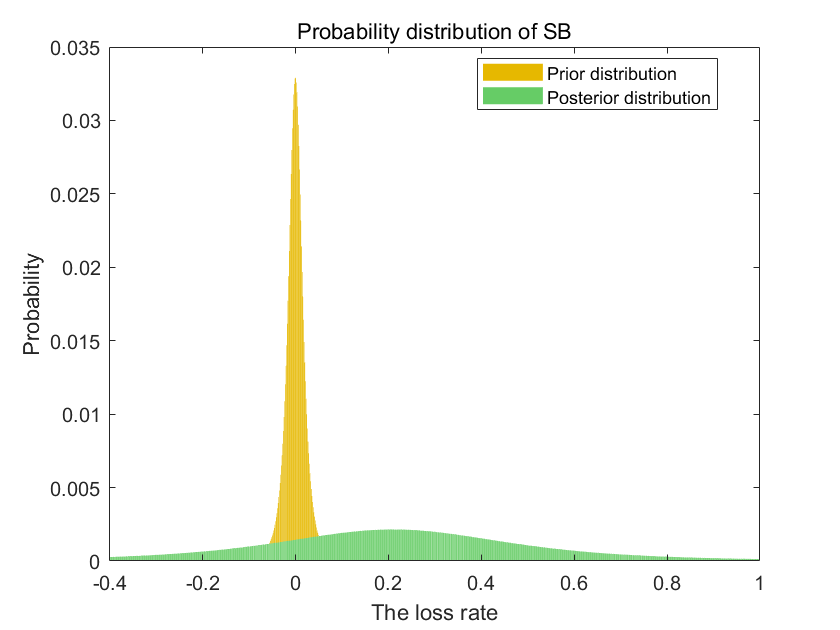}
    \caption{Prior and posterior distributions of loss for SB under the expectation view.}
  \end{subfigure}
   \begin{subfigure}[b]{0.45\textwidth}
    \centering
    \includegraphics[width=3.38 in]{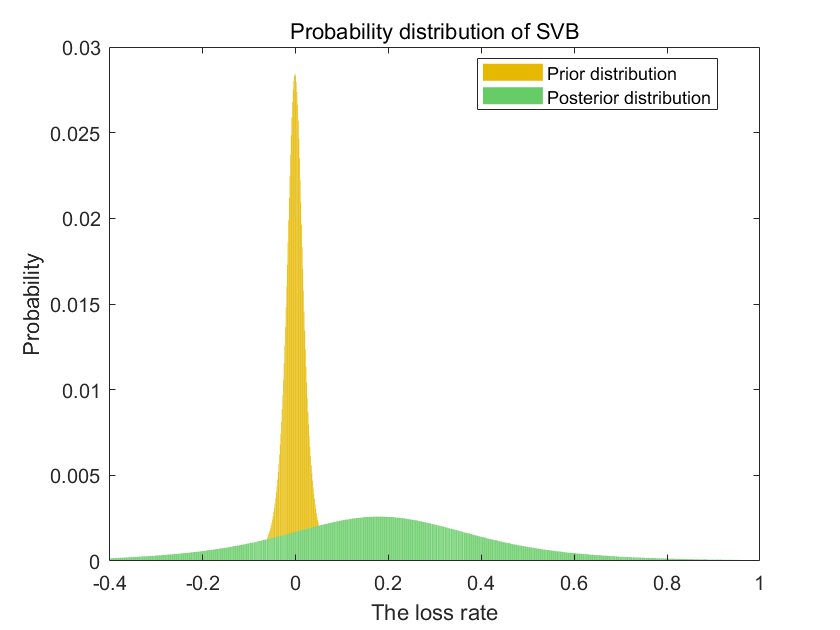}
    \caption{Prior and posterior distributions of loss for SVB under the quantile view.}
  \end{subfigure}
  \hfill
  \begin{subfigure}[b]{0.45\textwidth}
    \centering
    \includegraphics[width=3.38 in]{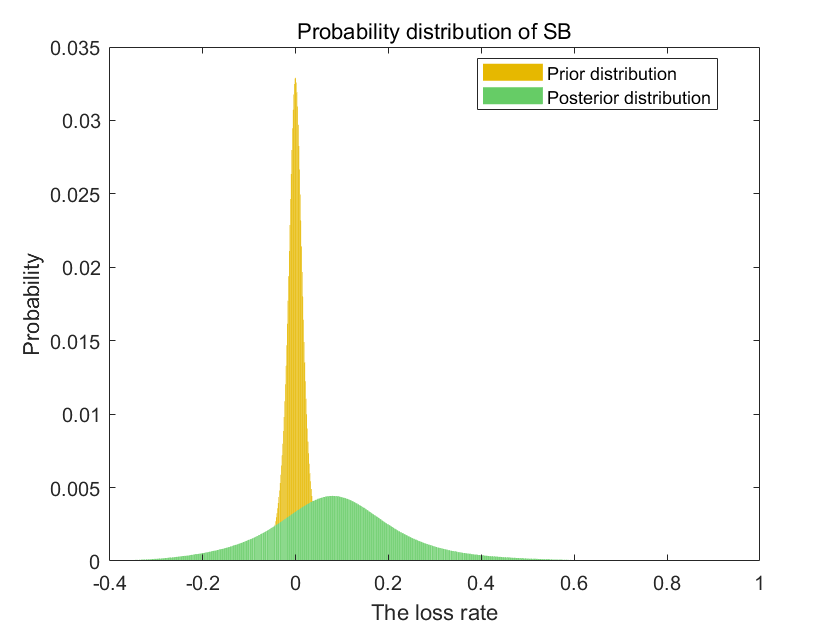}
    \caption{Prior and posterior distributions of loss for SB under the quantile view.}
  \end{subfigure}
\caption{Prior and posterior distributions of losses for SVB and SB under the expectation view and the quantile view.}
\label{distribution}
\end{figure}

To illustrate the mechanics of the EP method, we present the prior and posterior distributions of losses for SVB and SB under the expectation view and the quantile view on SVB's loss in Fig. \ref{distribution}.  These views on SVB lead to an increase in both the expectation and variance of SB's loss, while the distribution type remains unchanged.

To verify the robustness of our conclusions, we further estimate the general CoVaR under various views regarding FRB see Appendix \ref{CoVaR FRB}. The results indicate that the CoVaR estimates based on expectation-based views effectively cover the worst losses experienced by banks in the three days following FRB's failure.

\subsubsection{The general \texorpdfstring{$\Delta$}{Delta}\text{CoVaR} and risk spillover}

To evaluate SVB's risk spillover effects on other banks, we estimate the $\Delta$CoVaR for the five banks and the NBI.
The results are presented in Table \ref{delta CoVaR}. $\Delta$CoVaR is calculated as the difference between the general CoVaR under a specific view and its corresponding VaR, representing the additional risk induced by SVB under that particular view.

Our results indicate that under the expectation view (a), the magnitude of SVB's risk spillover effects on other banks and the NBI aligns with their CoVaR ranking.  The estimated $\Delta$CoVaR for the NBI exceeds that of any individual bank, under the joint expectation and variance view (b), as well as the quantile view (c), indicating that SVB's spillover effect on the NBI is the strongest in these scenarios. Notably, under views (a), (b), and (c), the spillover effects from SVB to the banks and the NBI are significant, typically surpassing their corresponding losses, thereby underscoring SVB's systemic impact during extreme market conditions.

Under the traditional CoVaR condition (view (d)), although SB and the NBI have similar CoVaR estimates, SVB's spillover effect on SB is only 0.84\%, approximately one-third of its effect on NBI.  Similarly, while PACW and WAL both have CoVaR estimates of 6.48\%,  and FRB and the NBI share estimates of 4.80\%, PACW's $\Delta$CoVaR is lower than WAL's, and FRB's $\Delta$CoVaR is lower than NBI's. These differences  are attributable to the relatively higher VaR values of SB, FRB, and PACW, implying that these banks were already exposed to higher levels of risk, which further contributes to their eventual failure or merger in 2023.

\begin{table}[htbp]
  \centering
  \caption{The general $\Delta$CoVaR of five banks and NBI under various views on SVB.  }
  \renewcommand\arraystretch{1.3}
     \begin{tabular}{llrrrrrr}
  \toprule
&Views  &SB  &FRB   &FHN &WAL &PACW &NBI\\
    \hline
Loss (\%) & /  &22.87  & 14.84 & 3.97 & 20.88 & 37.91  &4.64\\
 \midrule
 \multirow{4}{*}{$\Delta$CoVaR  (\%)} %&    (a)  $X=60.41\%$ &     66.08  & 43.82  & 77.28  & 61.46  & 66.08  & 42.84  \\
  &    (a)  $\tilde \mu_X=60.41\%$  &67.76 &52.50  & 65.80  & 55.58 & 57.54 & 47.32 \\
   &    (b) $\tilde \mu_X=60.41\%, \tilde \sigma_X^2=3.60\%$ & 70.70 &71.40 &71.40 &72.10  &70.56 & 73.08  \\
    &    (c)  $\tilde q_X=60.41\%$   &31.22 &27.02 &27.72 & 30.94 &31.08 &31.78\\
  &    (d) $X=q_X=4.24\%$    & 0.84 & 2.10   & 2.66  & 2.94  & 2.80  & 2.24 \\
   \bottomrule
    \end{tabular}%
  \label{delta CoVaR}%
\end{table}%

\section{Conclusion}
\label{Sec 5}

This article proposes a general CoVaR framework based on entropy pooling, enabling the assessment of the CoVaR for assets under various types of views.
 
First, we show the numerical implementation based on the EP approach to estimate the general CoVaR, addressing the limitations of the quantile regression method and GARCH model, which can only handle specific forms of views (such as $X=l$ and $X\geq l$). Second, under the assumption that the prior and posterior distributions of asset losses follow a bivariate normal distribution,  we derive the analytical expressions for general CoVaR. By analyzing the sensitivity of CoVaR to different view values, we find CoVaR exhibits a linear relationship with the expection and the difference of expections the variables in the view, while demonstrating nonlinear relationships with variance, quantiles, and the correlation coefficient. 
Finally, in the context of the Fed's rate hikes and the 2023 US banking crisis, we estimate the general CoVaR for banks to comprehensively evaluate systemic risk under various scenarios. Specifically, the CoVaR model under the distribution view regarding rate hikes is found to be effective and the most stable.  To further validate the reliability of our general CoVaR, we also examine the CoVaR under a broader range of scenarios and views in Appendix \ref{Appendix C}.  

In summary, our general CoVaR effectively integrates diverse expert views, providing early warnings for systemic risk.

% Bibliography

\appendix

\section*{Appendix}

\section{\textbf{Analytical expressions of the general \texorpdfstring{$\Delta$}{Delta}\text{CoVaR} under bivariate normal distributions}}
\label{Appendix B}

The analytical expressions for the general \textnormal{$\Delta$CoVaR} under views based on quantile and correlation coefficient,  as well as under a relative view of two variables given bivariate normal distributions, are presented as follows.
 
\subsection*{\textbf{A.1.  \texorpdfstring{$\Delta$}{Delta}\text{CoVaR} under a view on quantile}}

Under a quantile view  $\tilde q_X=q_1$,  our \textnormal{$\Delta$CoVaR} for $Y$ is expressed as:
 \begin{equation} 
 \begin{aligned} \Delta\textnormal{CoVaR}^{Y|\tilde q_X=q_1}_\alpha
 &=\left\{
\begin{aligned}
 \rho (q_1-q_X) \frac{\sigma_Y}{\sigma_X} +\{\tilde\sigma_Y-\left[\tilde\sigma_Y^2-(1-\rho^2)\sigma_Y^2\right]^\frac{1}{2}-(1-\rho)\sigma_Y\}\Phi^{-1}(\alpha), \quad \text{if} \ \rho \geq 0,\\
  \rho (q_1-q_X) \frac{\sigma_Y}{\sigma_X} +\{\tilde\sigma_Y+\left[\tilde\sigma_Y^2-(1-\rho^2)\sigma_Y^2\right]^\frac{1}{2}-(1-\rho)\sigma_Y\}\Phi^{-1}(\alpha), \quad \text{if} \ \rho<0,\\
%\mu_Y+  \sigma_Y\Phi^{-1}(\alpha), \quad \rho=0\\
\end{aligned} 
 \right.
 \end{aligned} 
 \end{equation} 
 where $\tilde \sigma_Y=\sigma_Y\{\frac{1+(1-\rho^2)c^2}{1+c^2}+\frac{\rho^2(\kappa+c)c\left[(\kappa+c)c+\sqrt{(\kappa+c)^2c^2+4(1+c^2)}\right]}{2(1+c^2)^2}\}^\frac{1}{2}$, $\kappa= \frac{q_1- q_X}{\sigma_X}$, and $c=\Phi^{-1}(\alpha)$. 
Particularly, when $X$ and $Y$ are unrelated, the risk spillover effect of the view is nonexistent. 

When $X$ and $Y$ are perfectly positively correlated, we have $\Delta\textnormal{CoVaR}^{Y|\tilde q_X=q_1}_\alpha= (q_1-q_X) \frac{\sigma_Y}{\sigma_X}$.
 If the posterior 95\% VaR of $X$ exceeds its prior $\textnormal{VaR}^X_\alpha$, i.e., $q_1 > q_X$, indicating an increase in risk for  $X$, then $\Delta\textnormal{CoVaR}^{Y|\tilde q_X=q_1}_\alpha$ is directly proportional to the volatility ratio $\sigma_Y / \sigma_X$.  This suggests that a higher volatility ratio amplifies the sensitivity of asset $Y$'s risk to changes in asset $X$'s risk.

\vspace{2mm}

 \subsection*{\textbf{A.2.  \texorpdfstring{$\Delta$}{Delta}\text{CoVaR} under a view on correlation coefficient}}

Under a correlation coefficient view $\tilde \rho=\rho_1$, the \textnormal{$\Delta$CoVaR} for $Y$ is:
 \begin{equation} \begin{aligned} \Delta\textnormal{CoVaR}^{Y|\tilde \rho=\rho_1}_\alpha= \sigma_Y\left[\left(\frac{1-\rho^2}{1-\rho\rho_1}\right)^\frac{1}{2}-1\right] \Phi^{-1}(\alpha),\end{aligned}  \end{equation} 
where $\rho$ and $\rho_1$ both lie in the interval $(-1, 1)$. Notably, when $\rho=0$, the risk spillover effect of the view is nonexistent.  
If $\rho>0$, $\Delta\textnormal{CoVaR}^{Y|\tilde \rho=\rho_1}_\alpha$  increases as  $\rho_1$ rises. Conversely, if $\rho<0$, $\Delta\textnormal{CoVaR}^{Y|\tilde \rho=\rho_1}_\alpha$  decreases as $\rho_1$ rises.

 \subsection*{\textbf{A.3.  \texorpdfstring{$\Delta$}{Delta}\text{CoVaR} under a relative view of two variables}}

Under the relative view, where the difference between $X$ and $Y$ follows $\mathcal{N}(d, s^2)$, the $\Delta$CoVaR for $Y$ is expressed as:
 \begin{align}
\Delta\textnormal{CoVaR}^{Y|(X-Y) \sim  \mathcal{N}(d, s^2)}_\alpha
&= \frac{(\mu_X -\mu_Y - d) \sigma_Y (\sigma_Y - \rho \sigma_X)}{\sigma_X^2 - 2 \rho \sigma_X \sigma_Y + \sigma_Y^2} \notag \\
&+\sigma_Y\left[\left(1+  \frac{\left[ s^2 - (\sigma_X^2 - 2 \rho \sigma_X \sigma_Y + \sigma_Y^2) \right] \left( \sigma_Y-  \rho \sigma_X \right)^2}{\left( \sigma_X^2 - 2 \rho \sigma_X \sigma_Y + \sigma_Y^2 \right)^2}\right)^{\frac{1}{2}}-1\right]\Phi^{-1}(\alpha).
\end{align} 
We find that if $\rho > \sigma_Y / \sigma_X$, $\Delta\textnormal{CoVaR}^{Y|(X-Y) \sim  \mathcal{N}(d, s^2)}_\alpha$ increases as $d$ increases. Conversely, if $\rho < \sigma_Y/\sigma_X$, $\Delta\textnormal{CoVaR}^{Y|(X-Y) \sim  \mathcal{N}(d, s^2)}_\alpha$ decreases as $d$ increases.

\section{Robustness of the general CoVaR}
\label{Appendix C}

To check the robustness of the general CoVaR, we estimate the CoVaR for the NBI under different rate hike scenarios, as well as for banks and the NBI under various views regarding FRB.

\subsection*{\textbf{B.1. The CoVaR of NBI under ten rate hikes}} 
\label{Robustness under views of interest hikes}

Table \ref{NBI_CoVaR} presents the CoVaR estimation for the NBI under the traditional CoVaR condition ($X=q_X$), using both the QR and EP methods, as well as CoVaR estimates under the expectation view ($X=\mathit{erh}$) and the distribution view, using the EP method. 
The results show that  CoVaR estimates under the expectation view and the distribution view have the lowest average absolute differences from actual losses. 
Notably, the stability of the CoVaR estimate under the distribution view is the highest, with the lowest variance in absolute differences. Since the distribution view effectively captures a sufficient amount of market information, the general CoVaR model based on this condition provides a reliable forecast of systemic financial risk.

\begin{table}[htbp]
\centering
\caption{The general CoVaR of NBI under ten-time interest hikes using QR and EP methods.}
    \renewcommand\arraystretch{1.2}
 \resizebox{\textwidth}{!}{
\begin{tabular}{ccrc@{\hspace{20pt}} c@{\hskip 2em} ccc}
        \toprule 
    \multirow{2}{*}{Time} & \multicolumn{1}{c}{\multirow{2}[0]{*}{Date}}     & \multicolumn{1}{c}{\multirow{2}[0]{*}{Loss  (\%)}}   & \multicolumn{1}{c}{CoVaR (\%): QR} & & \multicolumn{3}{c}{CoVaR (\%): EP}\\
   \cmidrule{6-8}   
 &     &   & \multicolumn{1}{c}{$X=q_X$}   & &\multicolumn{1}{c}{$X=q_X$}  & \multicolumn{1}{c}{$X=\mathit{erh}$} & \multicolumn{1}{c}{Distribution} \\
     \midrule
  1 & 03/17/2022   & -1.29   &3.6  &  & 3.4   &  3.1   & 3.1  \\
  2 & 05/05/2022    & 2.71   & 3.6   &  & 3.4   & 3.4    &3.4    \\
  3 & 06/16/2022   & 2.31    & 4.0  &  &4.0    &3.7     & 3.7  \\
  4 & 07/28/2022   & -2.47   &4.3    &  &  4.0  & 4.0    & 4.0  \\
  5 & 09/22/2022   & 4.50    & 4.6 &  & \textbf{4.3}   &\textbf{4.3}   & \textbf{4.0}  \\
  6 & 11/03/2022   & 0.38    &4.2  &  &4.6  &4.6     &4.6   \\
  7 & 12/15/2022   & 3.60   &4.2   &  & 4.0   &4.0     &4.0   \\
  8 & 02/02/2023    & -4.35  &3.4  &  & 3.4   &  3.4   & 3.4  \\
  9   & 03/23/2023  &  1.10  &3.4  &  &3.7    &   3.4  & 3.4    \\
  10  & 05/04/2023  &  6.47  &\textbf{3.6} &   &\textbf{3.7}   &\textbf{3.4}  & \textbf{3.4}  \\
%\hline
%  Kupiec test& &      &Accept & &Accept &Accept   &Accept  \\
\hline
\multirow{2}{*}{\makecell{Mean  of \\ absolute differences}}& &    	& \multirow{2}{*}{3.2} &  	& \multirow{2}{*}{3.2} 	& \multirow{2}{*}{3.1}  & \multirow{2}{*}{3.1} 		 \\
  &&&&&&\\
\multirow{2}{*}{\makecell{Variance of \\ absolute differences}} & &  	& \multirow{2}{*}{6.9}  &  &  \multirow{2}{*}{6.7} 	& \multirow{2}{*}{6.8}	& \multirow{2}{*}{6.6} \\
  &&&&&&\\
  \bottomrule  
  \end{tabular}%
 }

 \raggedright

  \vspace{1mm}
\addtabletext{\small{\quad\quad Notes: The term ``absolute difference" refers to the absolute value of the difference between the CoVaR estimate and the actual loss. }}
 \label{NBI_CoVaR}
 \end{table}

 \subsection*{\textbf{B.2. The CoVaR of banks under views regarding FRB}}
  \label{CoVaR FRB}

The failure of FRB triggered a ``domino effect", causing significant declines in the stock prices of First Horizon Bank (FHN), Western Alliance Bank (WAL), and PacWest Bank (PACW), with each bank's stock price dropping by over 30\% before May 4, 2023. In contrast, the NBI experienced a relatively smaller drop during the same period.

\begin{table}[htbp]
  \centering
  \caption{The general  CoVaR of three banks and NBI under various views on FRB. }
  \renewcommand\arraystretch{1.3}
    \begin{tabular}{llrrrr}
           \toprule
&Views  &FHN &WAL &PACW &NBI \\
    \hline
Loss (\%) &\ / &   0.17 & 1.83 & 10.64 & 2.08 \\
Worst loss (\%)& \ / &   33.16 & 38.45 & 50.62 & 5.29\\
[2mm]
\hline
\multirow{7}{*}{CoVaR (\%)}%& (a) $X=43.30\%$ &   51.98  & 51.14  & 55.90  & 27.62 &Accept\\
 &(a) Expectation view 
 $\tilde \mu_X=43.30\%$ &56.88  &61.92  &63.46&41.34  \\
 & \multirow{2}{*}{\makecell{(b) Expectation and variance view   \\ [1mm] \hspace{2mm}$\tilde \mu_X=43.30\%, \tilde \sigma_X^2=5.28\%$}} &  \multirow{2}{*}{67.24} &  \multirow{2}{*}{67.80} &  \multirow{2}{*}{73.12} &  \multirow{2}{*}{67.52}  \\
 
 & &&& \\
 &(c)  Quantile view $\tilde q_X=43.30\%$ & 28.60&29.72&31.96 & 15.16\\
& \multirow{2}{*}{\makecell{(d) Traditional CoVaR condition \\ [1mm] 
\hspace{2mm}$X=q_X= 2.98\%$ }} & \multirow{2}{*}{5.36} & \multirow{2}{*}{6.76} & \multirow{2}{*}{7.04} & \multirow{2}{*}{4.52}   \\ 
& &&&& \\
 &(e) No view  & 3.12 &3.68 &\textbf{3.82} &2.56   \\
 
    \bottomrule
    \end{tabular}%

  \vspace{1mm}
\addtabletext{\small{\quad\quad Notes: The ``loss" refers to the loss experienced by the corresponding bank on the day of FRB's failure, while the ``worst loss" denotes the largest single-day loss incurred by the corresponding bank within the three days following FRB's failure.}}
\label{FRB CoVaR Copula Fixed}
\end{table}

We estimate CoVaR under various views regarding FRB, as shown in Table \ref{FRB CoVaR Copula Fixed}. 
Similar to Section \ref{CoVaR SVB}, the value of $l$ in views (a), (b), and (c) is set to FRB's loss rate on the day before its collapse, which was 43.30\%. The parameter $\sigma_1^2$ represents the variance of the loss rates for the ten days prior to the collapse, estimated at 5.28\%, while the 95\% quantile $q_X$ is 2.98\%.
Overall, the CoVaR rankings (from highest to lowest) across various views are  (b), (a), (c), (d), and (e). 
Furthermore, except for PACW, the VaR estimates for the other banks exceeded their actual losses. 
The CoVaR estimates under expectation-based views effectively captured the worst losses the banks experienced within three days following the FRB's bankruptcy.  Across all views, PACW exhibited the highest risk and ultimately announced its merger with Bank of
California  in July 2023.  These results further underscore the role of CoVaR as an early warning indicator of systemic risk.

\end{document}